\newcommand{\PrePrintStyle}{2} 

\newcommand{\AddDraftMark}{0} 

\ifnum\PrePrintStyle=1
  \documentclass[prd,aps,amsmath,preprint,showpacs,superscriptaddress,nofootinbib,longbibliography,linenumbers,floatfix]{revtex4-2}

\else
  \documentclass[prd,aps,amsmath,reprint,showpacs,superscriptaddress,nofootinbib,longbibliography,floatfix]{revtex4-2}
\fi

\usepackage{graphicx}  
\usepackage{dcolumn}   
\usepackage{bm}        
\usepackage{xspace}
\usepackage{units}
\usepackage{placeins}
\usepackage{float}
\usepackage{verbatim}
\usepackage{afterpage}
\usepackage[flushleft]{threeparttable}
\usepackage{array,booktabs,makecell}
\usepackage[colorlinks=true,linkcolor=green,anchorcolor=red, urlcolor=Blue]{hyperref}
\usepackage[dvipsnames]{xcolor}

\ifnum\AddDraftMark=1
  \usepackage{draftwatermark} 
  \SetWatermarkText{~Draft}
  \SetWatermarkScale{12}
 \fi

\usepackage[normalem]{ulem}  

\begin{document}

\newcommand{\ie}{{\it i.e.}\xspace}
\newcommand{\eg}{{\it e.g.}\xspace}
\newcommand{\red}[1]{{\color{red}#1}}
\newcommand{\blue}[1]{{\color{blue}#1}}

\newcommand{\chisq}{\ensuremath{\chi^{2}}\xspace}
\newcommand{\chisqmin}{\ensuremath{\chi^{2}_{\textrm{min}}}\xspace}
\newcommand{\chisqpen}{\ensuremath{\chi^{2}_{\textrm{pen}}}\xspace}
\newcommand{\ndof}{\ensuremath{\mathrm{N_{DoF}}}\xspace}
\newcommand{\nbins}{\ensuremath{\mathrm{N_{bins}}}\xspace}
\newcommand{\units}[1]{\ensuremath{~\mbox{#1}}}

\newcommand{\pip}{\ensuremath{\pi^+\xspace}\xspace}
\newcommand{\pim}{\ensuremath{\pi^-\xspace}\xspace}
\newcommand{\pipm}{\ensuremath{\pi^\pm\xspace}\xspace}
\newcommand{\pin}{\ensuremath{\pi^0\xspace}\xspace}
\newcommand{\numu}{\ensuremath{\nu_\mu\xspace}\xspace}
\newcommand{\nue}{\ensuremath{\nu_e\xspace}\xspace}
\newcommand{\numubar}{\ensuremath{\bar{\nu}_\mu\xspace}\xspace}
\newcommand{\nuebar}{\ensuremath{\bar{\nu}_e\xspace}\xspace}

\newcommand{\ccpip}{$\nu_\mu$CC1$\pi^{+}$\xspace}
\newcommand{\ccnpip}{$\nu_\mu$CC$N\pi^{+}$\xspace}
\newcommand{\ccnpin}{$\nu_\mu$CC$N\pi^0$\xspace}
\newcommand{\ccapin}{$\bar{\nu}_\mu$CC$1\pi^0$\xspace}

\newcommand{\Pmu}{\ensuremath{p_\mu}\xspace}
\newcommand{\Emu}{\ensuremath{E_\mu}\xspace}
\newcommand{\Enu}{\ensuremath{E_\nu}\xspace}
\newcommand{\Enudef}{\ensuremath{E_\nu=E_\mu+q_0}\xspace}
\newcommand{\Tmu}{\ensuremath{\theta_\mu}\xspace}
\newcommand{\Kpi}{\ensuremath{T_\pi}\xspace}
\newcommand{\Tpi}{\ensuremath{\theta_\pi}\xspace}
\newcommand{\qq}{\ensuremath{Q^{2}}\xspace}
\newcommand{\qqdef}{\ensuremath{Q^{2}=2E_\nu(E_\mu-p_\mu\cos\theta_\mu)-M_\mu^2}\xspace}
\newcommand{\Ev}{\ensuremath{E_{\nu}}\xspace}
\newcommand{\Wrec}{\ensuremath{W_{\mathrm{rec}}}}
\newcommand{\Ma}{\ensuremath{M_{\textrm{A}}}\xspace}
\newcommand{\Eavail}{\ensuremath{E_{\textrm{avail}}}\xspace}
\newcommand{\q}{\ensuremath{q_{3}}\xspace}
\newcommand{\qdef}{\ensuremath{q_{3}}=\sqrt{Q^2+q_0^2}\xspace}

\newcommand{\Mares}{\ensuremath{M_{\mathrm{A}}^\mathrm{res}}\xspace}
\newcommand{\Normres}{\ensuremath{\mathrm{NormRes}}\xspace}
\newcommand{\nonresonepi}{NonRes$1\pi$\xspace}
\newcommand{\nonrestwopi}{NonRes$2\pi$\xspace}
\newcommand{\ThetaPi}{$\pi$-iso\xspace}

\newcommand{\GENIE}{GENIE\xspace}
\newcommand{\NOvA}{NOvA\xspace}
\newcommand{\minerva}{MINER$\nu$A\xspace}
\newcommand{\MINOS}{MINOS\xspace}
\newcommand{\MiniBooNE}{MiniBooNE\xspace}

\newcommand{\genie}{{\small GENIE}\xspace}

\newcommand{\addcitation}{{\bf\color{red} [ADD CITATION]}}

\newcommand{\revisionAdd}[1]{{\bf\color{blue} \protect #1}}
\newcommand{\revisionStrike}[1]{{\bf\color{blue}\protect \sout{#1}}}

\newcommand{\marvinAdd}[1]{{\color{RoyalPurple} \protect #1}}
\newcommand{\marvinStrike}[1]{{\color{RoyalPurple}\protect \sout{#1}}}
\newcommand{\marvinComment}[1]{{\noindent \color{RoyalPurple} \small [MA: #1]}}

\newcommand{\minerbaAdd}[1]{{\color{black} \protect #1}}
\newcommand{\minerbaStrike}[1]{{\color{black}\protect \sout{#1}}}
\newcommand{\minerbaComment}[1]{{\noindent \color{black} \small [MB: #1]}}

\newcommand{\rikAdd}[1]{{\color{ForestGreen} \protect #1}}
\newcommand{\rikStrike}[1]{{\color{ForestGreen}\protect \sout{#1}}}
\newcommand{\rikComment}[1]{{\noindent \color{ForestGreen} \small [RG: #1]}}

\newcommand{\albertoAdd}[1]{{\color{Brown} \protect #1}}
\newcommand{\albertoStrike}[1]{{\color{Brown}\protect \sout{#1}}}
\newcommand{\albertoComment}[1]{{\noindent \color{Brown} \small [AG: #1]}}



\title{Measurement of inclusive charged-current $\nu_\mu$ scattering on hydrocarbon at $\langle E_{\nu} \rangle \sim$ 6 GeV with low three-momentum transfer}


\newcommand{\Rutgers}{Rutgers, The State University of New Jersey, Piscataway, New Jersey 08854, USA}
\newcommand{\Hampton}{Hampton University, Dept. of Physics, Hampton, VA 23668, USA}
\newcommand{\Dortmund}{Institute of Physics, Dortmund University, 44221, Germany }
\newcommand{\Otterbein}{Dept. of Physics, Otterbein University, 1 South Grove Street, Westerville, OH, 43081 USA}
\newcommand{\JMU}{James Madison University, Harrisonburg, Virginia 22807, USA}
\newcommand{\Florida}{University of Florida, Dept. of Physics, Gainesville, FL 32611}
\newcommand{\UCIrvine}{Dept. of Physics and Astronomy, University of California, Irvine, Irvine, California 92697-4575, USA}
\newcommand{\CBPF}{Centro Brasileiro de Pesquisas F\'{i}sicas, Rua Dr. Xavier Sigaud 150, Urca, Rio de Janeiro, Rio de Janeiro, 22290-180, Brazil}
\newcommand{\PUCP}{Secci\'{o}n F\'{i}sica, Departamento de Ciencias, Pontificia Universidad Cat\'{o}lica del Per\'{u}, Apartado 1761, Lima, Per\'{u}}
\newcommand{\INRM}{Institute for Nuclear Research of the Russian Academy of Sciences, 117312 Moscow, Russia}
\newcommand{\Jlab}{Jefferson Lab, 12000 Jefferson Avenue, Newport News, VA 23606, USA}
\newcommand{\Pittsburgh}{Dept. of Physics and Astronomy, University of Pittsburgh, Pittsburgh, Pennsylvania 15260, USA}
\newcommand{\Guanajuato}{Campus Le\'{o}n y Campus Guanajuato, Universidad de Guanajuato, Lascurain de Retana No. 5, Colonia Centro, Guanajuato 36000, Guanajuato M\'{e}xico.}
\newcommand{\Athens}{Dept. of Physics, University of Athens, GR-15771 Athens, Greece}
\newcommand{\Tufts}{Physics Dept., Tufts University, Medford, Massachusetts 02155, USA}
\newcommand{\WM}{Dept. of Physics, College of William \& Mary, Williamsburg, Virginia 23187, USA}
\newcommand{\FNAL}{Fermi National Accelerator Laboratory, Batavia, Illinois 60510, USA}
\newcommand{\Purdue}{Dept. of Chemistry and Physics, Purdue University Calumet, Hammond, Indiana 46323, USA}
\newcommand{\MCLA}{Massachusetts College of Liberal Arts, 375 Church Street, North Adams, MA 01247}
\newcommand{\UMD}{Dept. of Physics, University of Minnesota -- Duluth, Duluth, Minnesota 55812, USA}
\newcommand{\Northwestern}{Northwestern University, Evanston, Illinois 60208}
\newcommand{\UNI}{Universidad Nacional de Ingenier\'{i}a, Apartado 31139, Lima, Per\'{u}}
\newcommand{\Rochester}{University of Rochester, Rochester, New York 14627 USA}
\newcommand{\Austin}{Dept. of Physics, University of Texas, 1 University Station, Austin, Texas 78712, USA}
\newcommand{\USM}{Departamento de F\'{i}sica, Universidad T\'{e}cnica Federico Santa Mar\'{i}a, Avenida Espa\~{n}a 1680 Casilla 110-V, Valpara\'{i}so, Chile}
\newcommand{\Geneva}{University of Geneva, 1211 Geneva 4, Switzerland}
\newcommand{\Chicago}{Enrico Fermi Institute, University of Chicago, Chicago, IL 60637 USA}
\newcommand{\hired}{}
\newcommand{\OregonState}{Dept. of Physics, Oregon State University, Corvallis, Oregon 97331, USA}
\newcommand{\oxford}{Oxford University, Dept. of Physics, Oxford, United Kingdom}
\newcommand{\umiss}{University of Mississippi, Oxford, Mississippi 38677, USA}
\newcommand{\upenn}{Dept. of Physics and Astronomy, University of Pennsylvania, Philadelphia, PA 19104}
\newcommand{\AMU}{AMU Campus, Aligarh, Uttar Pradesh 202001, India}
\newcommand{\wroclaw}{University of Wroclaw, plac Uniwersytecki 1, 50-137 Wrocław, Poland}
\newcommand{\Mohali}{IISER, Mohali, Knowledge city, Sector 81, Manauli PO 140306}
\newcommand{\ICL}{The Blackett Laboratory,  Imperial College London,  London SW7 2BW, United Kingdom}

\newcommand{\emilymaherThanks}{Department of Physics}
\newcommand{\mateusfcarneiroThanks}{Now at Brookhaven National Laboratory, Upton, New York 11973-5000, USA}
\newcommand{\mascencioThanks}{Now at Iowa State University, Ames, IA 50011, USA}
\newcommand{\byaeggyThanks}{Now at Department of Physics, University of Cincinnati, 
Cincinnati, Ohio 45221, USA}
\newcommand{\amitbashyalThanks}{Now at  High Energy Physics/Center for Computational Excellence Department, Argonne National Lab, 9700 S Cass Ave, Lemont, IL 60439}
\newcommand{\finerThanks}{Now at Los Alamos National Laboratory, Los Alamos, New Mexico 87545, USA}
\newcommand{\kleykampThanks}{Now at Department of Physics and Astronomy, University of Mississippi, Oxford, MS 38677}
\newcommand{\ricfregianThanks}{Now at Department of Physics and Astronomy, University of California, Davis, Davis,  CA 95616, USA}
\author{M. V.~Ascencio}\thanks{\mascencioThanks}   \affiliation{\PUCP}
\author{D.A. Andrade}                     \affiliation{\Guanajuato}
\author{I. Mahbub}                        \affiliation{\UMD}


\author{S.~Akhter}                        \affiliation{\AMU}
\author{Z.~~Ahmad~Dar}                    \affiliation{\WM}  \affiliation{\AMU}
\author{F.~Akbar}                         \affiliation{\Rochester} \affiliation{\AMU}
\author{V.~Ansari}                        \affiliation{\AMU}
\author{A.~Bashyal}\thanks{\amitbashyalThanks}                       \affiliation{\OregonState}
\author{S.~Bender}                        \affiliation{\UMD}
\author{A.~Bercellie}                     \affiliation{\Rochester}
\author{M.~Betancourt}                    \affiliation{\FNAL}
\author{A.~Bodek}                         \affiliation{\Rochester}
\author{J.~L.~Bonilla}                    \affiliation{\Guanajuato}
\author{K.~Bonin}                       \affiliation{\UMD}
\author{H.~Budd}                          \affiliation{\Rochester}
\author{G.~Caceres} \thanks{\ricfregianThanks}                       \affiliation{\CBPF}
\author{T.~Cai}                           \affiliation{\Rochester}
\author{M.F.~Carneiro}\thanks{\mateusfcarneiroThanks}  \affiliation{\OregonState}  \affiliation{\CBPF}
\author{G.A.~D\'{i}az~}                   \affiliation{\Rochester}
\author{H.~da~Motta}                      \affiliation{\CBPF}
\author{J.~Felix}                         \affiliation{\Guanajuato}
\author{L.~Fields}                        \affiliation{\FNAL}  \affiliation{\Northwestern}
\author{A.~Filkins}                       \affiliation{\WM}
\author{R.~Fine}\thanks{\finerThanks}     \affiliation{\Rochester}
\author{N.~Fuad}                           \affiliation{\UMD}
\author{A.M.~Gago}                        \affiliation{\PUCP}
\author{H.~Gallagher}                     \affiliation{\Tufts}
\author{P.K.Gaur}                         \affiliation{\AMU}
\author{A.~Ghosh}                         \affiliation{\USM}  \affiliation{\CBPF}
\author{R.~Gran}                          \affiliation{\UMD}
\author{T.~Haluptzok}                      \affiliation{\UMD}
\author{D. A.~Harris}                     \affiliation{\FNAL}
\author{S.~Henry}                         \affiliation{\Rochester}
\author{S.~Jena}                          \affiliation{\Mohali}
\author{D.~Jena}                          \affiliation{\FNAL}
\author{J.~Kleykamp}\thanks{\kleykampThanks}                      \affiliation{\Rochester}
\author{A.~Klustová}                      \affiliation{\ICL}
\author{M.~Kordosky}                      \affiliation{\WM}
\author{D.~Last}                          \affiliation{\upenn}
\author{A.~Lozano}                        \affiliation{\CBPF}
\author{X.-G.~Lu}                         \affiliation{\oxford}
\author{E.~Maher}                         \affiliation{\MCLA}
\author{S.~Manly}                         \affiliation{\Rochester}
\author{W. A.~Mann}                       \affiliation{\Tufts}
\author{C.~Mauger}                        \affiliation{\upenn}
\author{K. S.~McFarland}                  \affiliation{\Rochester}  \affiliation{\FNAL}
\author{J.~Miller}                        \affiliation{\USM}
\author{J. G.~Morf\'{i}n}                 \affiliation{\FNAL}
\author{J. K.~Nelson}                     \affiliation{\WM}
\author{C.~Nguyen}                        \affiliation{\Florida}
\author{A.~Olivier}                       \affiliation{\Rochester}
\author{V.~Paolone}                       \affiliation{\Pittsburgh}
\author{G. N.~Perdue}                     \affiliation{\FNAL}  \affiliation{\Rochester}
\author{K.-J.~Plows}                      \affiliation{\oxford}
\author{M.A.~Ram\'{i}rez}                 \affiliation{\upenn}  \affiliation{\Guanajuato}
\author{H.~Ray}                           \affiliation{\Florida}
\author{B.J.~Reed}                           \affiliation{\UMD}
\author{P.A.~Rodrigues}                   \affiliation{\umiss}  \affiliation{\Rochester}
\author{D.~Ruterbories}                   \affiliation{\Rochester}
\author{M.~Sajjad~Athar}                  \affiliation{\AMU}
\author{H.~Schellman}                     \affiliation{\OregonState}  \affiliation{\Northwestern}
\author{C. J.~Solano~Salinas}             \affiliation{\UNI}
\author{H.~Su}                            \affiliation{\Pittsburgh}
\author{M.~Sultana}                       \affiliation{\Rochester}
\author{E.~Valencia}                      \affiliation{\WM}  \affiliation{\Guanajuato}
\author{N.H.~Vaughan}                     \affiliation{\OregonState}
\author{A.V.~Waldron}                     \affiliation{\ICL}
\author{C.~Wret}                          \affiliation{\Rochester}
\author{B.~Yaeggy}\thanks{\byaeggyThanks}  \affiliation{\USM}
\author{K.~Yang}                          \affiliation{\oxford}
\author{L.~Zazueta}                       \affiliation{\WM}

\collaboration{The \minerva Collaboration} 
\noaffiliation

\date{\today}
%
%
\begin{abstract}
The \minerva experiment reports double-differential cross-section measurements for
$\nu_{\mu}$-carbon interactions with three-momentum transfer $|\vec{q}| < 1.2$ GeV obtained with medium energy
exposures in the NuMI beam.
These measurements are performed as a function of the three-momentum transfer and an energy transfer estimator called the available energy defined as the energy that would be visible in the detector.  The double differential cross sections are compared to the GENIE and NuWro predictions along with the modified version of GENIE which incorporates new models for better agreement with earlier measurements from \minerva.  In these measurements, the quasi-elastic, resonance, and multi-nucleon knockout processes appear at different kinematics in this two-dimensional space.  The results can be used to improve models for neutrino interactions needed by neutrino oscillation experiments.
\end{abstract}

\maketitle

%
%
\section{Introduction}
Neutrino oscillation physics is evolving to a precision era. One of the field's central goals is to measure the leptonic CP-phase, which, via leptogenesis, could help explain the baryonic asymmetry in the universe. 

Among other factors \cite{DUNE:2015lol,Hyper-KamiokandeProto-:2015xww,Acero:2019ksn,Abe:2018wpn}
the precision measurements of the CP-phase and the neutrino oscillation parameters rely on improvements to neutrino detection technology, enlargement of neutrino detector volumes and increases in beam intensities.

Measuring the leptonic CP-phase will also demand improved knowledge of the neutrino scattering cross sections involved. Such knowledge, indispensable for extracting the neutrino oscillation parameters,  is needed in order to obtain the correct reconstructed energy of the interacting neutrino from the observed final state information, and to predict backgrounds. 

The dependence on the nuclear environment of neutrino-scattering cross sections, which is relevant for scattering from targets heavier than hydrogen or deuterium, is challenging to model.  This is especially true for processes at low three-momentum
transfer such as quasielastic (QE) and $\Delta(1232)$ resonance production, and any processes that fill in the kinematic ``dip region" between the two.  These are the most important fraction of the events at the current accelerator-based long-baseline neutrino oscillation experiments  performing \cite{Acero:2019ksn,Abe:2018wpn} (or planning to perform) the measurements above. Thus, there is an urgent need to compare new developments in nuclear modeling with updated measurements of the low three-momentum transfer charged-current 
cross section.

This work reports a new measurement of the low three-momentum transfer charged-current double-differential cross-section made by the  \minerva experiment, using data from the medium energy beam configuration. The data used in this analysis correspond to an exposure of 10.61$\times{10}^{20}$ protons on target with a peak neutrino energy of approximately 6 GeV.

These results expand upon the original result with the low-energy data \cite{Rodrigues:2015hik}, which demonstrated the need for multi-nucleon effects such as a two-particle two-hole (2p2h) process and a cross-section suppression at the lowest energy transfer like the one provided by a RPA screening effect.  This new result provides higher statistics at the original three-momentum transfer, expands the range in three-momentum transfer to 1.2~GeV, incorporates both empirical and theoretical improvements to modeling since the original publication, and incorporates improvements to the 
neutrino flux prediction and detector simulation.

We compare newly available models of nuclear effects through this new measurement.  Some model alternatives can be implemented within our full detector simulation directly or via reweights.   These include the Super-scaling (SuSA) 2p2h model \cite{Dolan:2019bxf}, two updates of the Rein Sehgal resonance model \cite{Kabirnezhad:2017xzx, Berger:2007rq}, a low-$Q^2$ suppression of resonances such as in \cite{Stowell:2019zsh,Adamson:2014pgc} or the application of a nucleon removal energy cost to resonance events, and an enhancement of QE events in the dip region motivated by nucleon initial state treatments like a spectral function.  As will be shown in Sec.~\ref{sec:simulation}, they can be switched on individually or in small groups and compared directly to reconstructed data distributions to gain physics insight specific to a single process.  This analysis also informs choices for a better central value and uncertainties to use for the unfolding procedure compared to the previous result. 

In Sec.~\ref{sec:results}, the cross section is also compared to predictions from GENIE version 3 and NuWro. The unfolded cross section allows for a wider range of comparisons and can be used by researchers without requiring the full \minerva experiment simulation. This comes with a cost of modest model biases and increased uncertainties after unfolding.  The process of unfolding from measured hadronic energy in the detector to true energy available to make a signal excludes neutron and other missing energy from unbinding nucleons.  This observable is easily predicted by neutrino event generators and does not incur model uncertainties as would a traditional quantity like energy transfer.

%
%
\section{Experiment}
The \minerva detector was situated on axis in the NuMI  beamline at Fermilab. The active region consists of 208 hexagonal planes made of triangular plastic scintillator strips. The detector is segmented longitudinally into several subdetectors: nuclear targets, active scintillator tracker, and downstream electromagnetic and hadronic calorimeters (ECAL and HCAL) ~\cite{Aliaga:2013uqz}. There is an HCAL-like outer subdetector that radially surrounds all inner detector. 
The tracker is made solely of scintillator planes. Each hexagonal plane of the tracker is composed of 127 nested polystyrene scintillator strips of varying length,  
with a triangular cross section of base 3.3 cm, and height 1.7 cm. The target mass of the fiducial volume is a mix of carbon (88.51\%), hydrogen (8.18\%), oxygen (2.5\%), titanium (0.47\%), chlorine (0.2\%), aluminum (0.07\%), and silicon (0.07\%). The planes alternate between three orientations, $0^\circ$ and $\pm 60^\circ$ with respect to the vertical, allowing an accurate three-dimensional reconstruction ~\cite{Aliaga:2013uqz} of the interaction point and muon track angle, even when hadronic activity 
overlaps with energy deposited by 
the muon.  Wavelength-shifting fibers embedded in the strips of scintillator are read out by optical cables that are connected to photomultiplier tubes. The photomultiplier tubes read out the scintillation light and achieve 3~ns timing resolution.

The MINOS Near Detector, situated two meters downstream of the \minerva detector, served as a magnetized muon spectrometer~\cite{Michael:2008bc}. Muons tracks which exit the downstream end of \minerva are matched to tracks in the MINOS Near Detector, which provides a measurement of the muon charge and momentum.

The neutrino flux for the data presented here peaks at 6~GeV and contains 95\%~\numu, and contamination
consisting of \numubar, \nue, and \nuebar~\cite{MINERvA:2019hhc}. The neutrino beam is simulated with  G{\small EANT}4 9.2.p03~\cite{AGOSTINELLI2003250} and constrained with thin-target hadron production measurements \cite{Aliaga:2016oaz} and an {\em in-situ} neutrino electron scattering constraint~\cite{MINERvA:2019hhc}.

%
%
\section{Event Selection}
An inclusive charged-current \numu interaction sample is selected from events in \minerva's 5.3-ton fiducial volume in the medium energy beam.   Reconstructed muon tracks beginning in the fiducial volume of  \minerva are matched to corresponding tracks in the MINOS 
~\cite{Michael:2008bc}.
The muon momentum is calculated by using the ionization energy loss for a muon traversing the material in the  \minerva detector in conjunction with the momentum reconstructed from MINOS ~\cite{Michael:2008bc}. Muon charge is reconstructed using track curvature in the magnetized MINOS Near Detector.
  The energy estimated in MINOS is added to \minerva's muon energy estimate to calculate \Emu and \Pmu. The muon track in \minerva from the neutrino interaction point is used to determine the muon angle \Tmu with respect to the beam direction. The event selection requires $\Tmu<20^\circ$ and $1.5 <\Pmu< 20.0$ \unit{GeV} to ensure good muon acceptance, which yielded a sample of 3,390,718 events with 98.64$\%$ purity. 

There are three main ingredients needed to reconstruct the kinematics of every event in this analysis: the hadronic energy deposition from all subdetectors in \minerva, the muon angle, and the muon energy. For the hadronic energy, a calorimetric correction derived from the simulation is applied following the same procedure used in the Low Energy (LE) beam measurements~\cite{Rodrigues:2015hik,Gran:2018fxa} to first obtain an estimator for the total lab-frame energy transfer $q_0$.  This is the same as other \minerva publications of inclusive cross sections and is described in
~\cite{Norrick:2018mwj}.
The reconstructed calorimetric neutrino energy is $E_{\nu}=E_{\mu}+q_0$ which is used to determine the four-momentum transfer $\qqdef$ and three-momentum transfer $|\vec{q}|$ or $\qdef$. The measurement region is limited to  $\q<$ \unit[1.2]{GeV}, which is a \unit[0.4]{GeV} expansion from the LE measurement.

The available energy is defined as 
\begin{equation}
    E_\text{avail}=\sum T_\text{proton}+\sum T_{\pi^{\pm}} +\sum E_\text{particle}
\end{equation}
where $T_\text{proton}$ is the proton kinetic energy, and $T_\pi$ is the charged pion kinetic energy, and $E_\text{particle}$ is the total energy of any other final state particles except neutrons. The definition excludes a nucleon mass from strange baryons. The reconstructed quantity input for $E_\text{avail}$ comes only from hadronic system activity in the tracker and ECAL regions of the \minerva detector, which improves the resolution at these lower hadronic energies. For unfolding we want a truth-level quantity that is measured by this reconstructed quantity. The choice for $E_\text{avail}$ is the same as energy transfer.  It does not account for neutron kinetic energy or missing energy from unbinding nucleons.   These produce little to no energy in the detector. The quantity $E_\text{avail}$, defined like this, minimizes model dependence while still being easy for neutrino event generators to predict.

\begin{figure}
\centering
\includegraphics[width=0.56\textwidth]{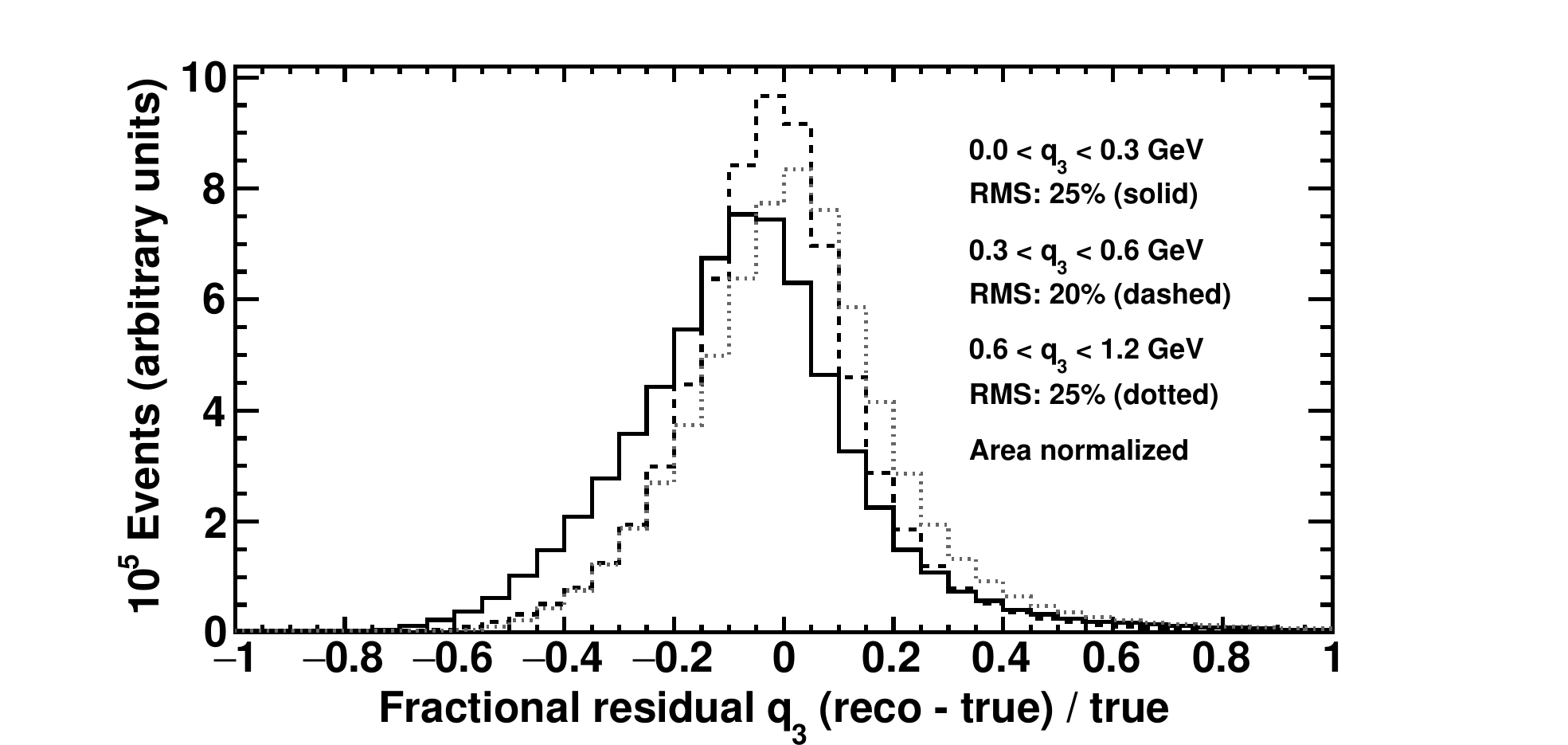}\\
\includegraphics[width=0.56\textwidth]{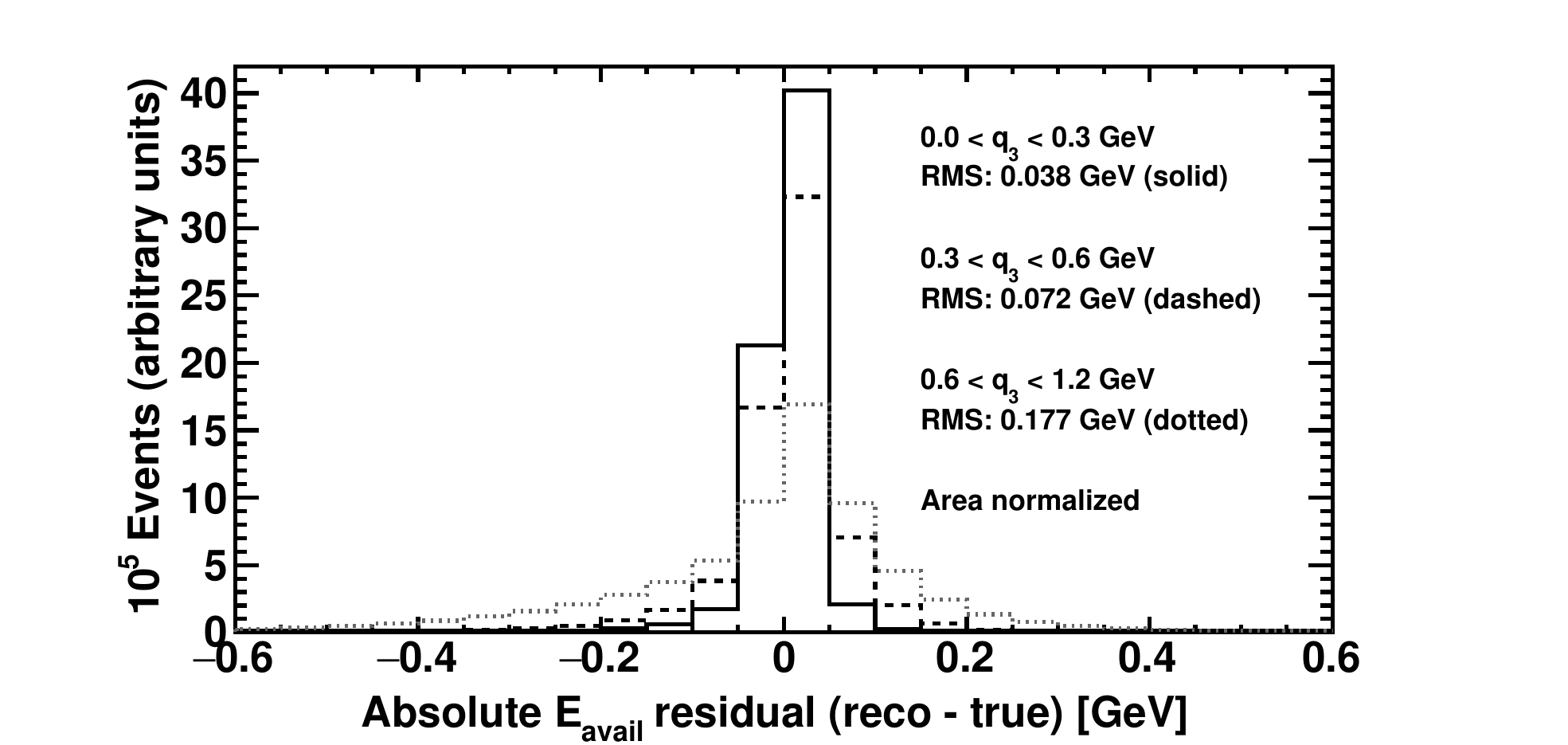}

\caption{Fractional resolution for three-momentum transfer $q_3$ (reconstructed-true)/true (top), and absolute resolution for $E_\text{avail}$ (reconstructed - true) in three regions of $q_3$ (bottom), for Monte Carlo simulated events passing the event selection with reconstructed $q_3 < 1.2$ \unit{GeV} .  The three different $q_3$ range samples are normalized to the same total area. }
\label{fig:Resolutions}
\end{figure}

Fig.~\ref{fig:Resolutions} shows the fractional resolution for $q_3$ and the absolute resolution for $E_\text{avail}$. The fractional resolution for $q_3$ is similar for the three regions with RMS ranging between 20\% or 25\%.  The resolution is driven mostly by the hadronic energy resolution, half as much by the muon angle resolution, and half again as much by the muon energy resolution.  The hadronic energy in $q_3$ includes activity in the outer detector and HCAL regions. It needs to be an estimator for energy transfer as described above and also cover a range of $q_3$ beyond the analysis sample.
  
This produces additional fluctuations not present in the $E_\text{avail}$ tracker + ECAL estimator . 
The muon angular resolution is better for small angles and degrades at larger angles. $E_\text{avail}$ is presented as an absolute quantity because the fractional resolution includes some events with a denominator close to zero.  For higher values of $q_3$, the calorimetry of interacting hadrons is an important part of the total resolution.   In the range $0.3 \text{ GeV} < q_3 < 0.6$ GeV, the resolution is expected to be that of the lowest energy stopping protons from our detector beam test \cite{MINERvA:2015yej}.  The central peak width of the resolution at the lowest momentum transfer is still proton-like.   In all three, the enhanced, negative side tail is partly from protons that G{\small EANT}4 turns into neutrons early in the detector leading to an underestimate of the true \Eavail .

\section{Models simulated compared to reconstructed data}
\label{sec:simulation}
The neutrino event generator GENIE~v.2.12.6 \cite{Andreopoulos:2009rq} is used to simulate neutrino interactions. The QE scattering model uses a relativistic Fermi gas model (RFG) incorporating a high-momentum tail in the Fermi momentum distribution \cite{PhysRevD.24.1400}, and uses a dipole axial form factor with $M_\text{A}^{\text{QE}}$ = 0.99 GeV \cite{Bradford:2006yz}. Resonant production is modeled using the Rein-Sehgal model \cite{Rein:1980wg}, with $M_\text{A}^{\text{res}}$ = 1.12 GeV.  The deep inelastic scattering (DIS) process is modeled using the 2003 Bodek-Yang model  \cite{Bodek:2002ps}. The hadron final states from DIS interactions are produced by the AGKY hadronization model \cite{Yang:2009zx}, which is tuned to reproduce hadron multiplicity data from $\nu$ and $\bar{\nu}$ bubble chamber experiments.  It uses Koba-Nielsen-Olsen scaling \cite{Koba:1972ng} below invariant mass of $W = 2.3$~GeV and transitions to use only {\small PYTHIA}  \cite{Sjostrand:2006za} for $W > 3.0$~GeV.  Of importance for the sample in this paper, the resulting DIS model in the tuned, low-$W$ regime provides the non-resonant background in the resonance region.
 
We use a version of GENIE 
modified with  \minerva-specific changes which we refer to as MnvTune-v1.2.   We use the Valencia RPA suppression \cite{Nieves:2004wx,Gran:2017psn} applied as a weight to QE events\cite{Gran:2017psn}, a non-resonant pion weight \cite{Rodrigues:2016xjj, Wilkinson:2014yfa} based on reanalysis of bubble chamber data \cite{Radecky:1981fn, Kitagaki:1986ct}, and a suppression of coherent production of pions with kinetic energy below 450~MeV based on  \minerva data \cite{MINERvA:2017ipy}.   We 
simulate 2p2h events using the Valencia 2p2h model \cite{Nieves:2011pp, Gran:2013kda, Schwehr:2016pvn}.  These were already part of the model used in \cite{Rodrigues:2015hik}.  To better describe 

the data in \minerva's previous low recoil result, the 2p2h event rate is enhanced in the kinematic region between QE and $\Delta$ reactions.   This enhancement was obtained by fitting a two-dimensional Gaussian function in true energy and momentum transfer to the neutrino data in \cite{Rodrigues:2015hik}.  The tune to neutrino data was applied and first described in the anti-neutrino analysis \cite{Gran:2018fxa}, then later to most of \minerva analyses.  Finally, two more weights are added to correct errors in the \genie FSI elastic scattering and pion absorption models \cite{Harewood:2019rzy}.

In addition to \genie changes, improvements to the flux constraints, reconstruction, and detector modeling also affect the comparison with data compared to the previous result.   The ab-initio neutrino flux \cite{Aliaga:2016oaz} is reduced by 12\% based on a measurement of neutrinos scattering off atomic electrons \cite{MINERvA:2019hhc}. The fully-simulated reduction of the detector mass by 2\% is based on the measurement of two as-built scintillator planes.  The muon energy from the MINOS Near Detector portion of muon tracks in data is scaled by +3.6\% following a study of the dependence of the event rate as a function of $E_\nu$ distribution \cite{MINERvA:2021mpk}.   The rest of the detector simulation 
uses {\small GEANT4} \cite{AGOSTINELLI2003250} version 9.4.p02 with uncertainties based on \minerva's hadron test beam data \cite{MINERvA:2015yej} and external data constraints \cite{Ashery:1981tq,Allardyce:1973ce, Saunders:1996ic,Lee:2002eq,Abfalterer:2001gw,Schimmerling:1973bb, Slypen:1995fm, Franz:1989cf,Tippawan:2008xk, Bevilacqua:2013rfq}.   The simulation of this medium-energy beam configuration includes the changing beam intensity throughout the era in both the beam structure of the simulated events, 
and the simulation of accidental beam-related activity
that affects the reconstruction.  A small, data-driven correction to the muon matching efficiency between  \minerva and MINOS Near Detector is applied as a function of instantaneous proton beam intensity.  

\begin{figure*}[ht!]
\centering
\includegraphics[width=0.55\textwidth]{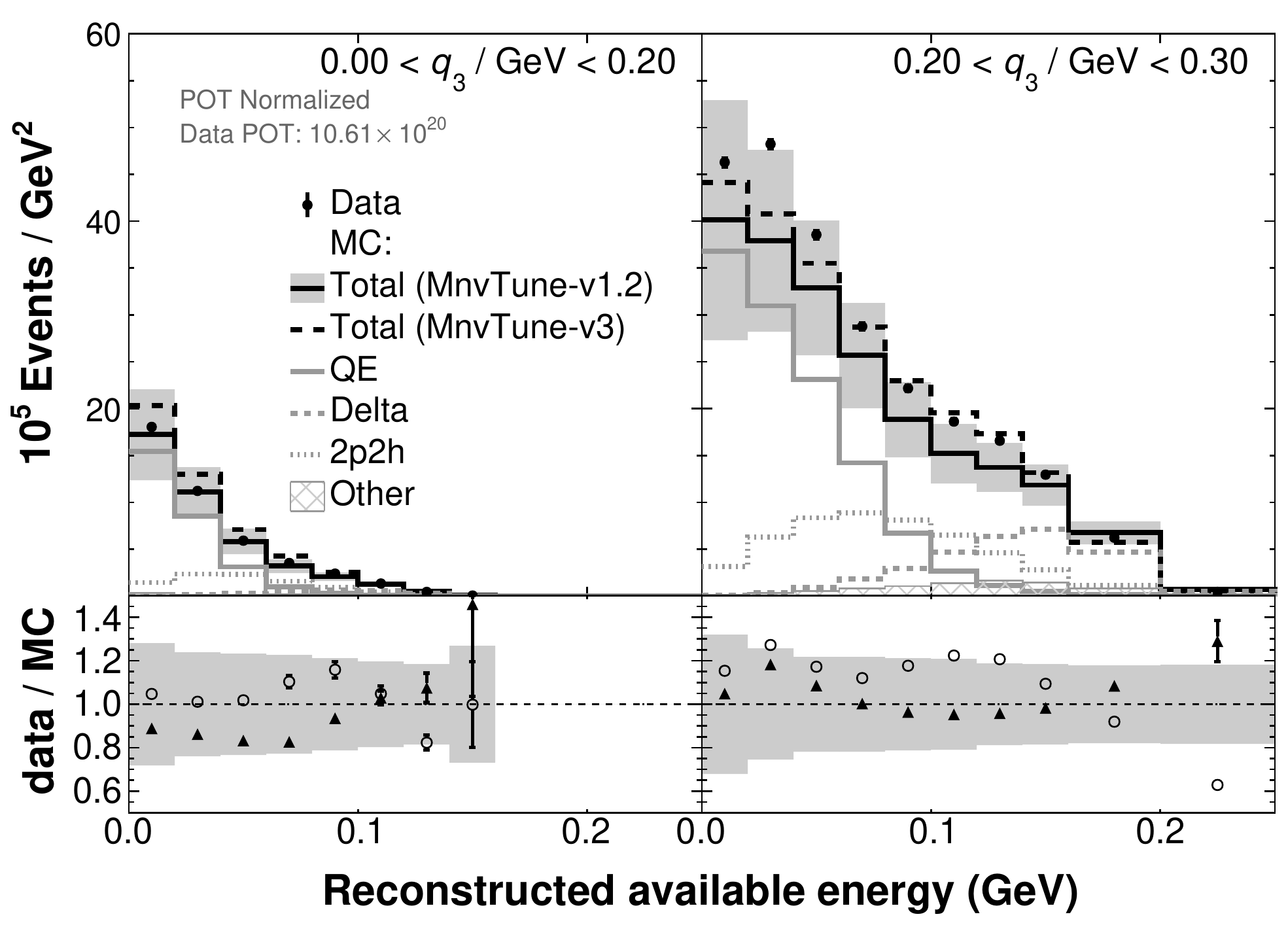}\\
\includegraphics[width=0.55\textwidth]{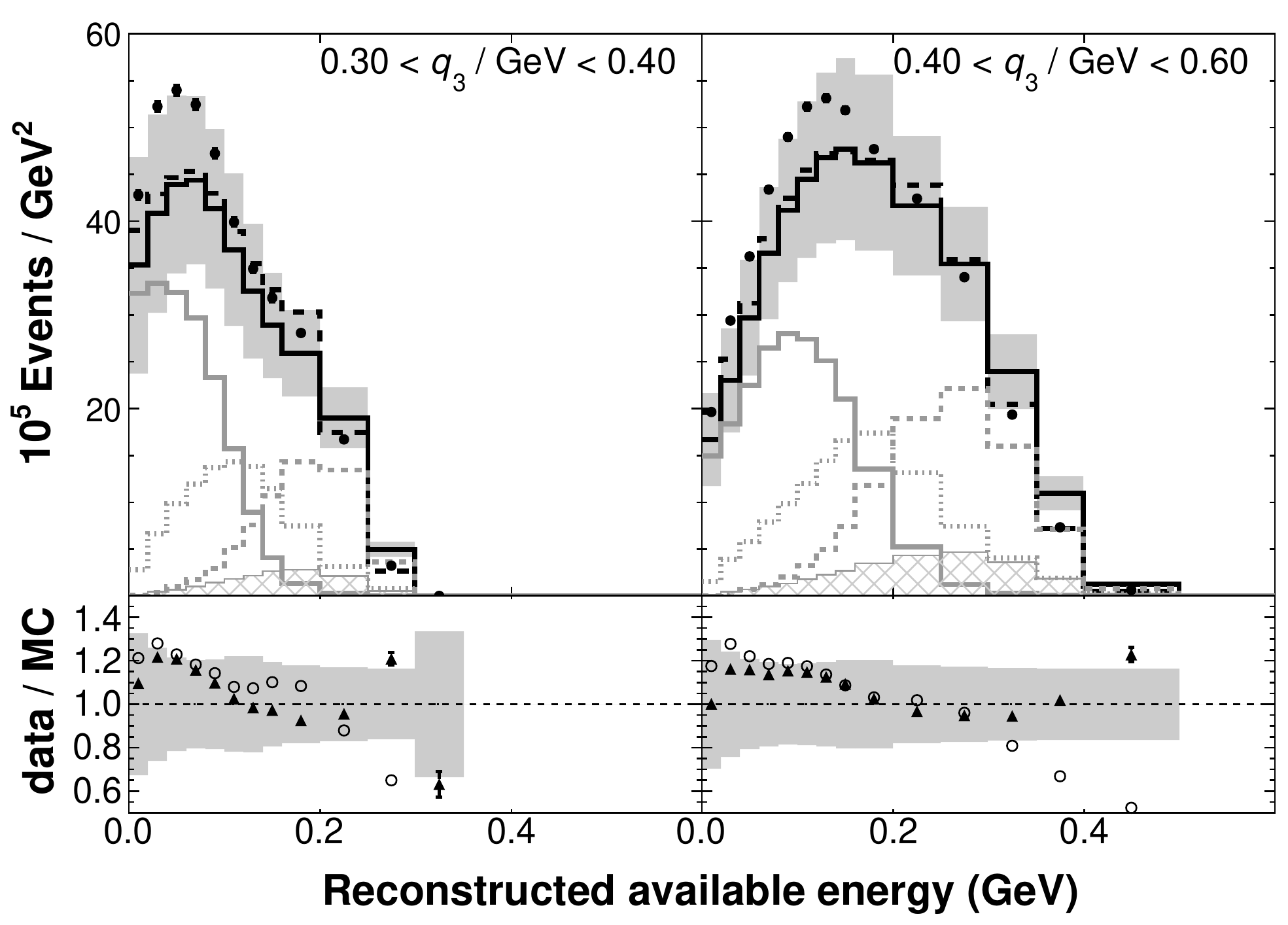}\\
\includegraphics[width=0.55\textwidth]{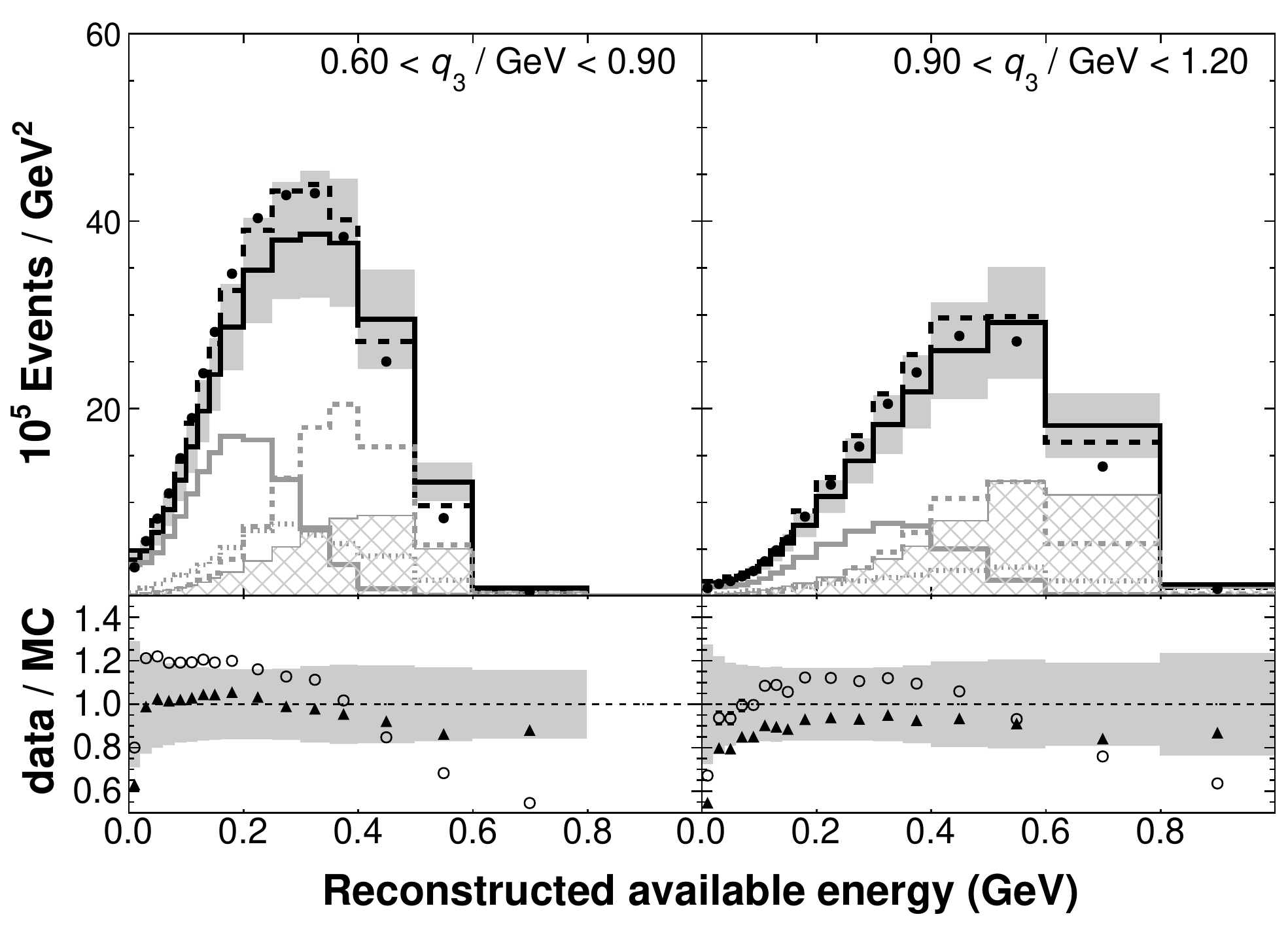}
\caption{Event rate as a function of reconstructed three-momentum transfer and available energy. Points with statistical uncertainties only are shown in data and for most points are too small to see. The simulation (MC) is the MnvTune-v1.2 (solid black) configuration and includes the breakdown into the predicted QE with RPA screening (solid gray), $\Delta$ resonance (dashed gray) with no low-$Q^2$ suppression, the tuned Valencia 2p2h contribution (dotted gray), and an Other category that is the mix of resonances and non-resonance background in the resonance region (gray grid filled histogram).  The uncertainty band includes all the interaction model and experimental systematic uncertainties.  A MnvTune-v3 prediction (dashed black) is used as the central value for unfolding and is described in the text. The ratio plots correspond to data divided by MnvTune-v1.2 (circles) and MnvTune-v3 simulation (triangles). 
}
\label{fig:lowRecoilMnv}
\end{figure*}

The resulting data and prediction in the two-dimensional kinematic space of reconstructed three-momentum transfer and available energy show areas of disagreement.    Shown in Fig.~\ref{fig:lowRecoilMnv} with the solid black line, the QE (solid gray line), $\Delta$ (dashed gray line), and 2p2h (dotted gray line) subcomponents provide most of the event rate, with another (gray grid-filled histogram) category that is a mix of higher resonances, non-resonant pion production in the resonance region and above, and coherent pion production.  The error band includes all the standard experimental and interaction model uncertainties used for earlier \minerva data,  as discussed in detail in Section VI.  

\subsection{Additional model variations}
To better understand the model elements that could describe these data, some of the latest models are incorporated into a further analysis of the reconstructed distributions in Fig.~\ref{fig:lowRecoilMnv}.  The tuned (MnvTune-v1.2) simulation does not describe the data well, and it contains a 2p2h model that was empirically tuned to prior \minerva data \cite{Rodrigues:2015hik}.  The prediction 
is improved with a low-$Q^2$ resonance modification such as in \cite{Stowell:2019zsh}, but that also is an empirical modification based on  \minerva data, which is similar to the one measured by MINOS using neutrino scattering on iron \cite{Adamson:2014pgc}.  

By incorporating the latest theoretical work into the analysis of Fig.~\ref{fig:lowRecoilMnv} it is clear why a new tuning is needed. The result is the so-called MnvTune-v3 (dashed line) of the default \genie v2.12.
Compared to MnvTune-v1.2, MnvTune-v3 changes each of the three main processes in one way.   The Valencia 2p2h model and its enhancement are replaced with the SuSA 2p2h model \cite{RuizSimo:2016rtu,Megias:2016fjk,Dolan:2019bxf}, which has more events in the dip region and fewer at very low $Q^2$.  The QE process has the initial nucleon's high momentum tail enhanced, which increases the QE rate overall and especially outside the QE peak.  And the outgoing hadronic system for resonances has 25 MeV removed from events with at least one proton in the final state; this especially moves $\Delta$ events into the dip region and away from very low $Q^2$.  These are not the only options and the rest of the section describes the effort in detail.

Most data are within the error band for MnvTune-v1.2, and the model would be in better agreement if a low-$Q^2$ pion suppression were added.  The MnvTune-v3 model shows a better agreement for most but not all regions of this kinematic space. That leads to a better unfolded cross section and the related studies improve our confidence in the uncertainty estimates.  An oscillation experiment using these tunes and uncertainties could expect good systematic uncertainty coverage.  The absolute discrepancy and the uncertain underlying physics mechanism would remain concerns.  If the interaction model uncertainties are not considered, the data are well outside the  \minerva experiment's energy scale and flux uncertainties, which combine to be 10\% for most bins.  These data will serve as an important benchmark for future models, whose ultimate validity can be confirmed at
this level of precision. 

\subsection{The 2p2h and QE region}

The MnvTune-v1.2 is compared to a new prediction based on the SuSA implementation of a 2p2h model \cite{RuizSimo:2016rtu,Megias:2016fjk,Dolan:2019bxf}  and an enhancement of the \genie QE high-momentum ``Bodek-Ritchie" tail \cite{PhysRevD.23.1070}.  These changes are a core element of MnvTune-v3.  This section describes the evolution of the models from the \genie default to MnvTune-v3, and the theory motivations for the modified 2p2h model followed by the QE model.

\subsubsection{Tuned Valencia and SuSA 2p2h models}

The MnvTune-v1.2 was motivated by the original data and analysis \cite{Rodrigues:2015hik}: its prediction was missing strength in the region between the QE and $\Delta$ peaks in the event rate, hereafter called the ``dip" region.  That analysis already used the screening effect known as RPA \cite{Nieves:2004wx} added to the generated QE events.  Additional events were simulated according to the Valencia 2p2h model \cite{Nieves:2011pp, Gran:2013kda, Schwehr:2016pvn}.  Additional missing strength was both absolute and relative to the QE and $\Delta$ regions.  Using a fit, we extracted a scale factor for the 2p2h component to match the data only in the dip region.  The fit function is a two-dimensional Gaussian of true $q_3$ and $q_0$ with parameters that allow the fit to locate the enhancement, align to the dip region, and determine a strength that describes the data.  Fit only to neutrino data in \cite{Rodrigues:2015hik}, the result was first deployed in \cite{Gran:2018fxa, MINERvA:2018vjb, MINERvA:2018hqn, MINERvA:2018hba}.    This single neutrino-mode tune improves the description of a number of  \minerva observables in both neutrino and anti-neutrino data, lepton and hadron quantities.   The tune comes with two alternate forms that enhance only the proton-neutron ($pn$) and not-$pn$ initial states and are used for uncertainty studies.    Compared to the \genie default, the RPA effect and this 2p2h scaling are the most significant components of MnvTune-v1.2 in the kinematic region of this analysis.

The dip region is predicted to be roughly equal mix of QE, 2p2h, and $\Delta$ resonances, yet the tune described in the previous paragraph only modifies the 2p2h component.   A third alternate 2D Gaussian fit that enhances only the QE process does a poor job of describing the data, especially non-calorimetric distributions from the transverse kinematic imbalance analysis \cite{MINERvA:2018hba}.  This led to a different strategy to probe the QE content in the dip region, described later in this section.   Theory motivated ways to enhance the $\Delta$ production in the dip region are described in the next section.

Since that effort, a new 2p2h model has become available for use by experiments.  The SuSA collaboration needed a 2p2h model \cite{RuizSimo:2016rtu,Megias:2016fjk,Dolan:2019bxf} to accompany their SuSA-motivated mean-field model for the QE \cite{Megias:2016fjk, Amaro:2019zos} in order to describe neutrino interactions without pions in the final state.  Because both the Valencia and SuSA 2p2h implementations in \genie for carbon use the same framework \cite{Schwehr:2016pvn}, 
the fully simulated Valencia 2p2h events in \minerva 's \genie 2.12 can be reweighted to represent the SuSA 2p2h prediction.  The results from this reweight reproduces the original SuSA \genie 3.0 based prediction nearly perfectly in energy and momentum transfer and produces only a small shift of about 10 MeV in the hadronic energy.

The SuSA 2p2h model inherently puts more events into the dip region.  In the Valencia 2p2h model, there is a population of events which explicitly have $\Delta$ kinematics.  In the SuSA model, this population appears at lower invariant mass, and therefore peaks at lower hadronic energy.  Thus, it is able to account for part of the missing strength in the dip region that required the original 2p2h tune.  It also significantly reduces the prediction at high $q_3$ but low $Q^2$, has more cross section strength averaged over all kinematics, and has intrinsically more events with two protons in the final state, before intranuclear rescattering. 

In order to extend the SuSA 2p2h prediction to its full $q_3 < 2.0$ GeV range,
the Valencia model was regenerated in that same range.  Because the Valencia model has a non-relativistic component in the calculation, it produces an unphysical prediction especially at high $q_3$ but low $Q^2$.  When not weighting to SuSA, an additional weight reduces the prediction to zero in that region, 
 keeping the Valencia prediction without the unphysical events. In addition to enabling the full range of the SuSA prediction, this analysis needs the extended range for an estimate of migration effects around the  1.2~GeV cutoff of the original model.

\subsubsection{QE events from high momentum nucleons}

In a Fermi-gas model, the QE process is approximately symmetric around its energy transfer peak at a given $q_3$, with a spread coming from the initial momentum of the struck nucleon.  There are a number of theoretical approaches that describe a tail to this distribution at higher energy transfer that will populate the dip region.  Some extract this feature from electron scattering data \cite{Amaro:2004bs}.   For others it arises from a nuclear mean field \cite{Pandey:2014tza, Megias:2016fjk, Amaro:2019zos}.  We seek to mimic these features using the fully simulated and reconstructed MC events available in this analysis. The SuSA QE prediction has been made available in \genie 3.0.6 \cite{Dolan:2019bxf}, in a way that, in the context of this work, we can only use it to reweight muon kinematics, hence we do not apply it here. 

Another way to enhance the QE events in the dip region is to explicitly enhance events where the struck nucleon had unusually high momentum.   The simulated \genie prediction is a relativistic fermi gas with a tail added from 221 MeV to 500 MeV of initial nucleon momentum according to the prescription of Bodek and Ritchie \cite{PhysRevD.24.1400}.  This tail is understood to come from the presence of short range correlated pairs of nucleons.  However, the population simulated when \genie uses this prescription is much lower than is determined by \cite{Benhar:1994hw,Benhar:2005dj} for example.  \minerva's implementation weights up MnvTune-v1.2 QE events by a factor of 6 at 221~MeV and reduces the weight to a factor of 1 (i.e. no change) at 500~MeV or below 221~MeV, increasing the QE rate by 24\%.   Because the event rate in Fig.~\ref{fig:lowRecoilMnv} is underpredicted, we do not weight down the rest of the QE population to compensate, but we continue to apply the RPA suppression.  Overall the predicted QE rate is 7\% higher than having neither weight.   A similar strategy is independently introduced in a recent update to \genie \cite{GENIE:2021npt}.  As in \genie, the implied spectator nucleon is not simulated.

\begin{figure*}[ht!]
\centering
\includegraphics[width=0.65\textwidth]{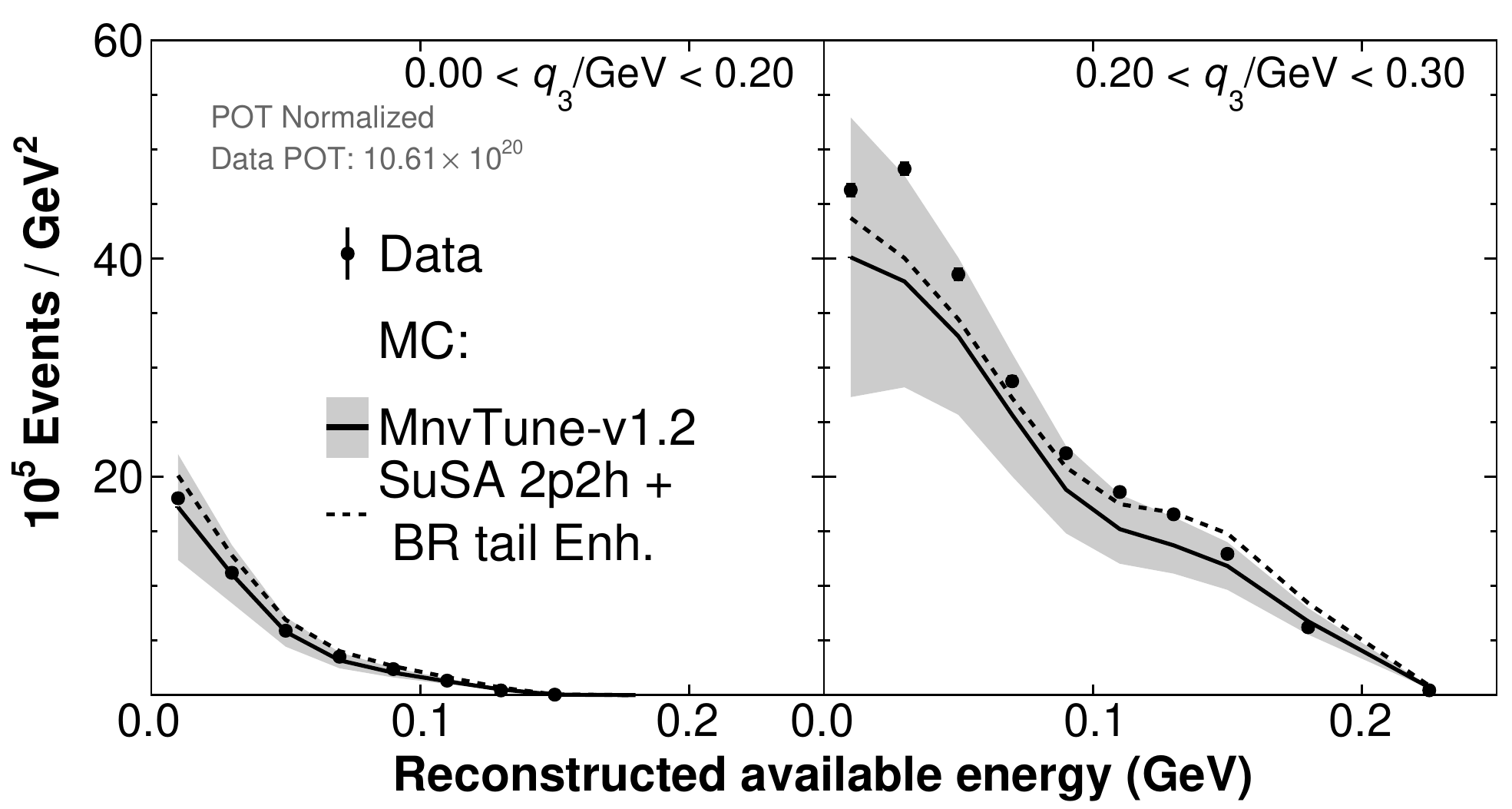}\\
\includegraphics[width=0.65\textwidth]{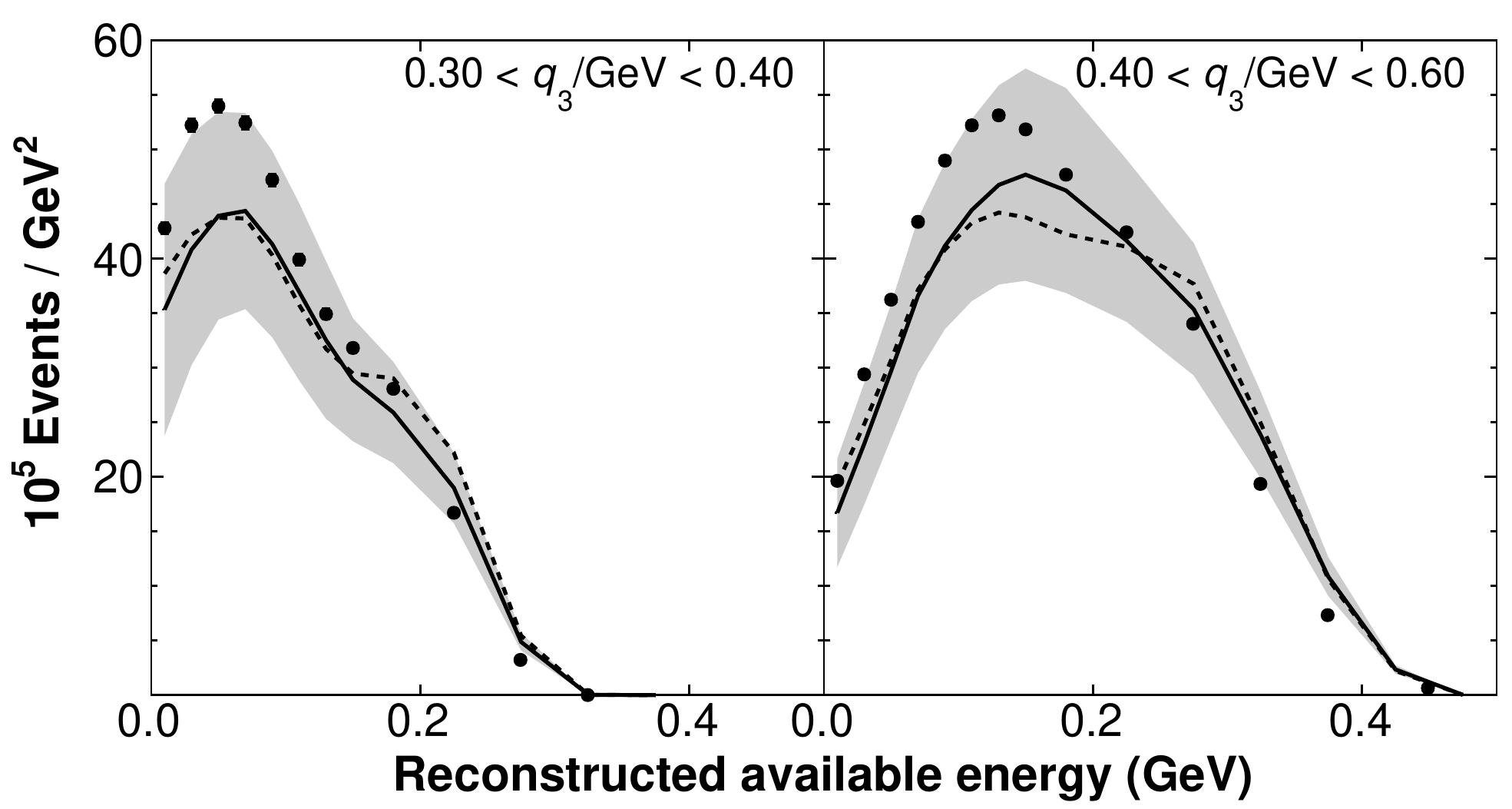}\\
\includegraphics[width=0.65\textwidth]{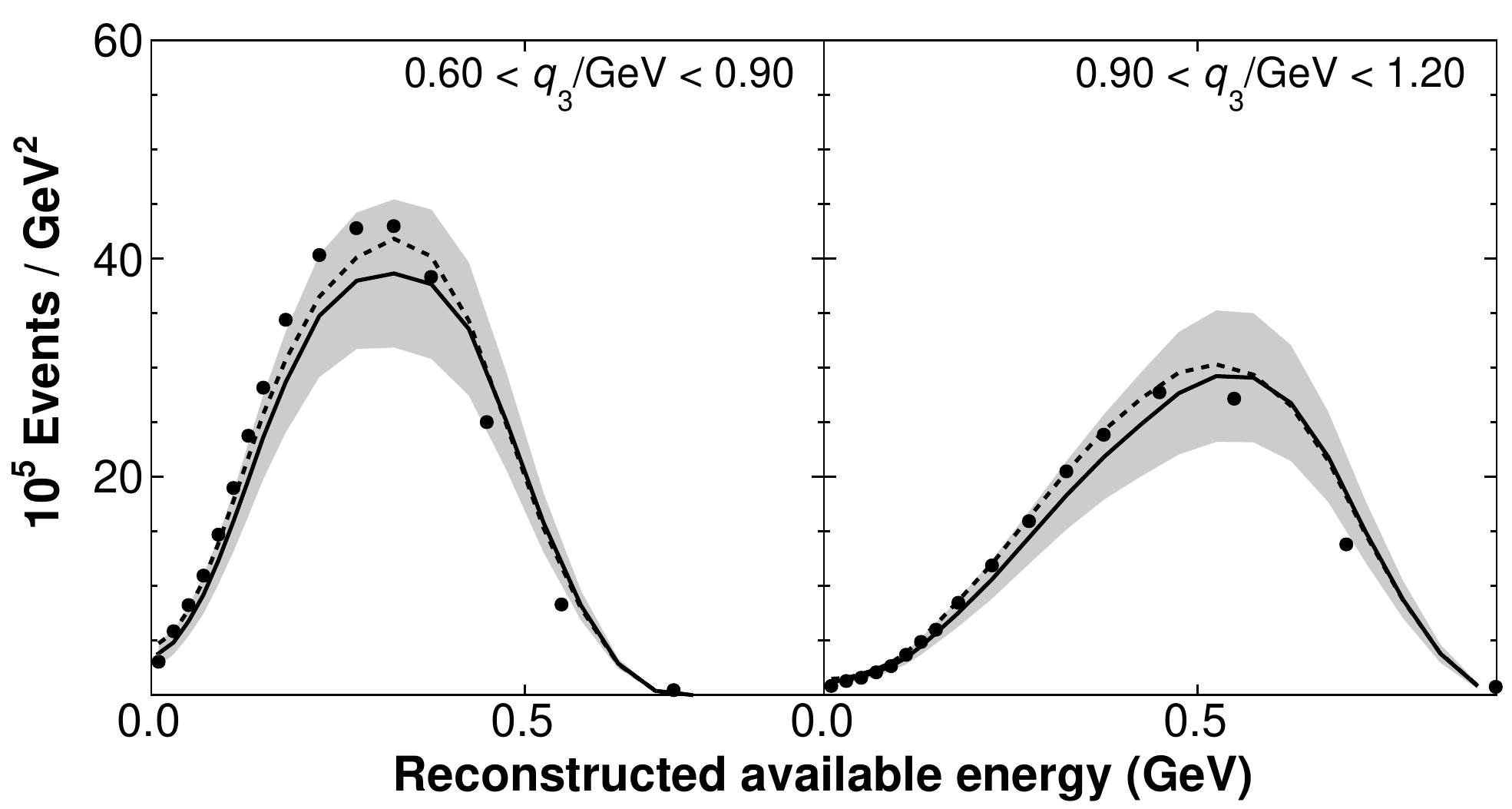}
\caption{Comparison of the MnvTune-v1.2 (solid black) and SuSA-2p2h+QE-tail-enhancement (dashed black) and the  \minerva reconstructed distributions (Points with Statistical Uncertainties only are shown in data).  The two models have similar ability to describe the QE and into the region between the QE and $\Delta$ resonance.
}
\label{fig:lowRecoilMECQEstudy}
\end{figure*}

The comparison is shown in Fig.~\ref{fig:lowRecoilMECQEstudy} with the original QE and tuned 2p2h (solid black) to the new QE with an enhanced tail and the SuSA 2p2h prediction (dashed black), but no changes to the resonance events.  Overall the theory-motivated configuration achieves a similar prediction to the MnvTune-v1.2 and better describes the data for some but not all kinematics.   

The descriptions of 2p2h and QE are still not as accurate as desired. Adjusting parameters, $M_\text{A}^{\text{QE}}$ or the strength of the RPA effect in the current models is possible. Further improvement to the axial form factor could be done incorporating the lattice QCD results \cite{kronfeld2019,meyer2022status}. Of more interest are new approaches for calculating the 2p2h and QE processes, including the hadronic systems, that are becoming available for analysis of data. They include full spectral functions which combines a realistic description of the initial target state with a fully relativistic interaction vertex and kinematics\cite{Benhar:1994hw}, the short time approximation which accounts reliably for two-nucleon dynamics, including correlations and currents, and provides information on back-to-back nucleons\cite{pastori}.

\subsection{Resonance Model Studies}

\begin{figure*}[ht!]
\centering
\includegraphics[scale=0.65]{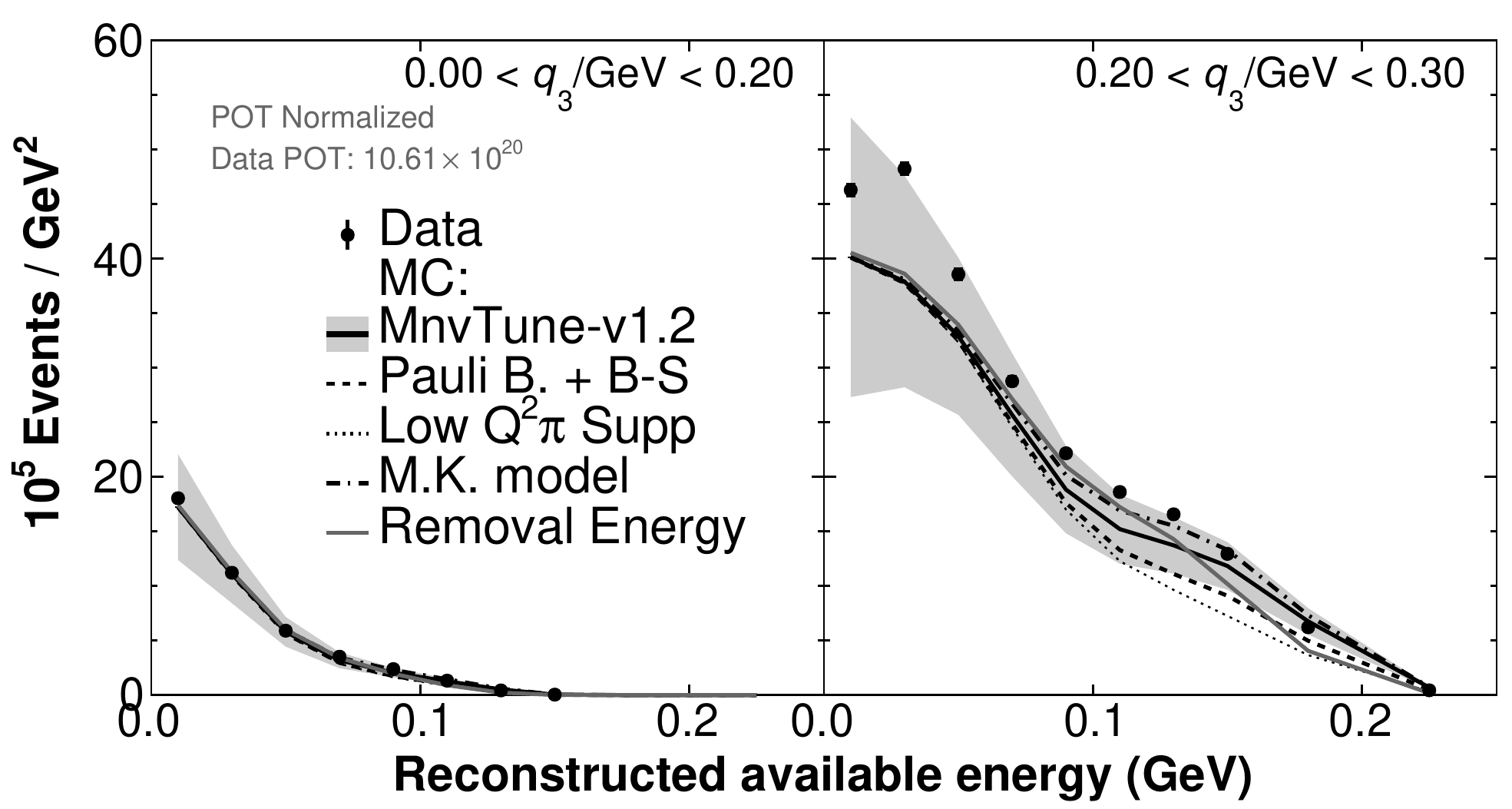}\\
\includegraphics[scale=0.65]{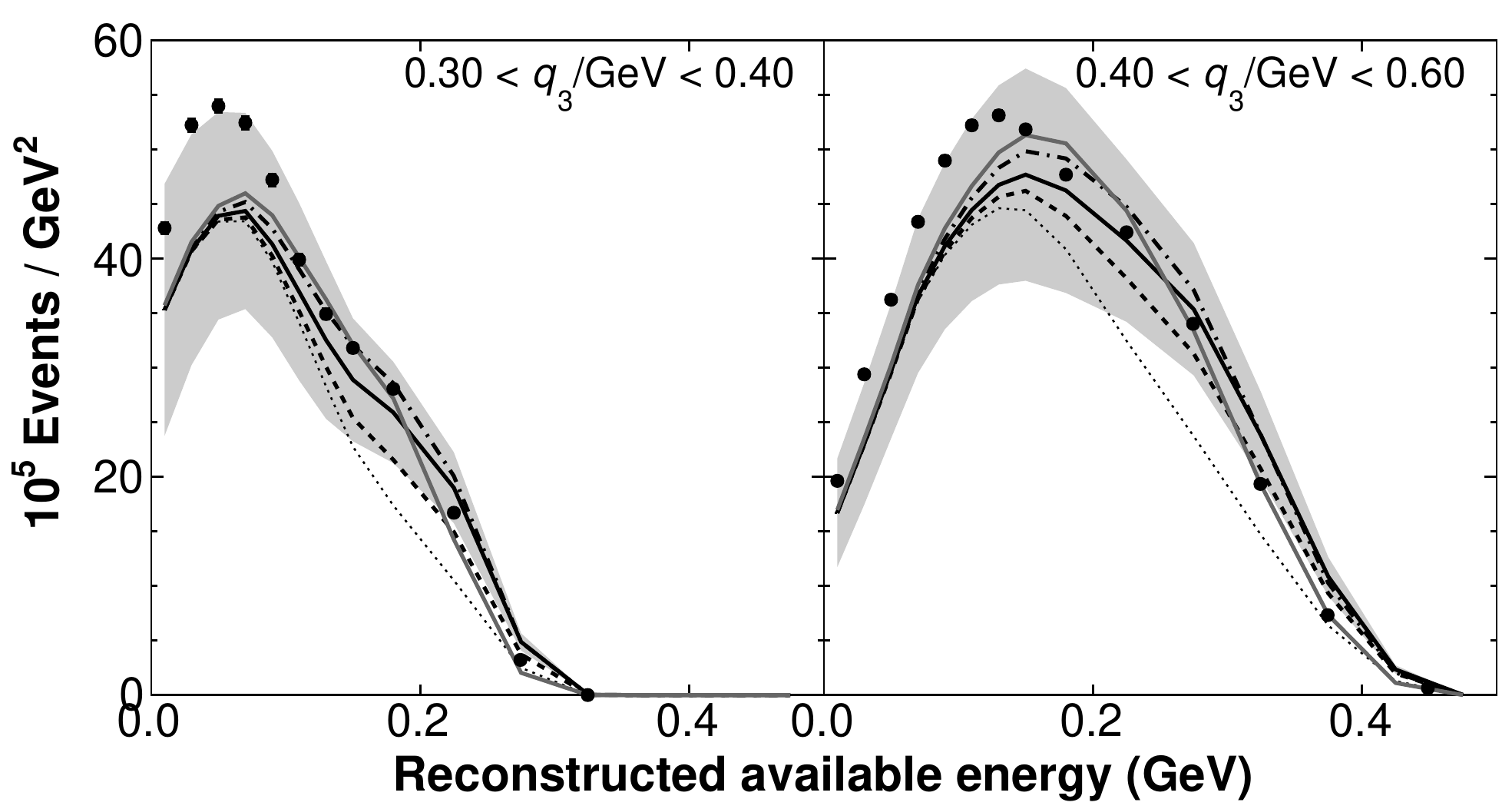}\\
\includegraphics[scale=0.65]{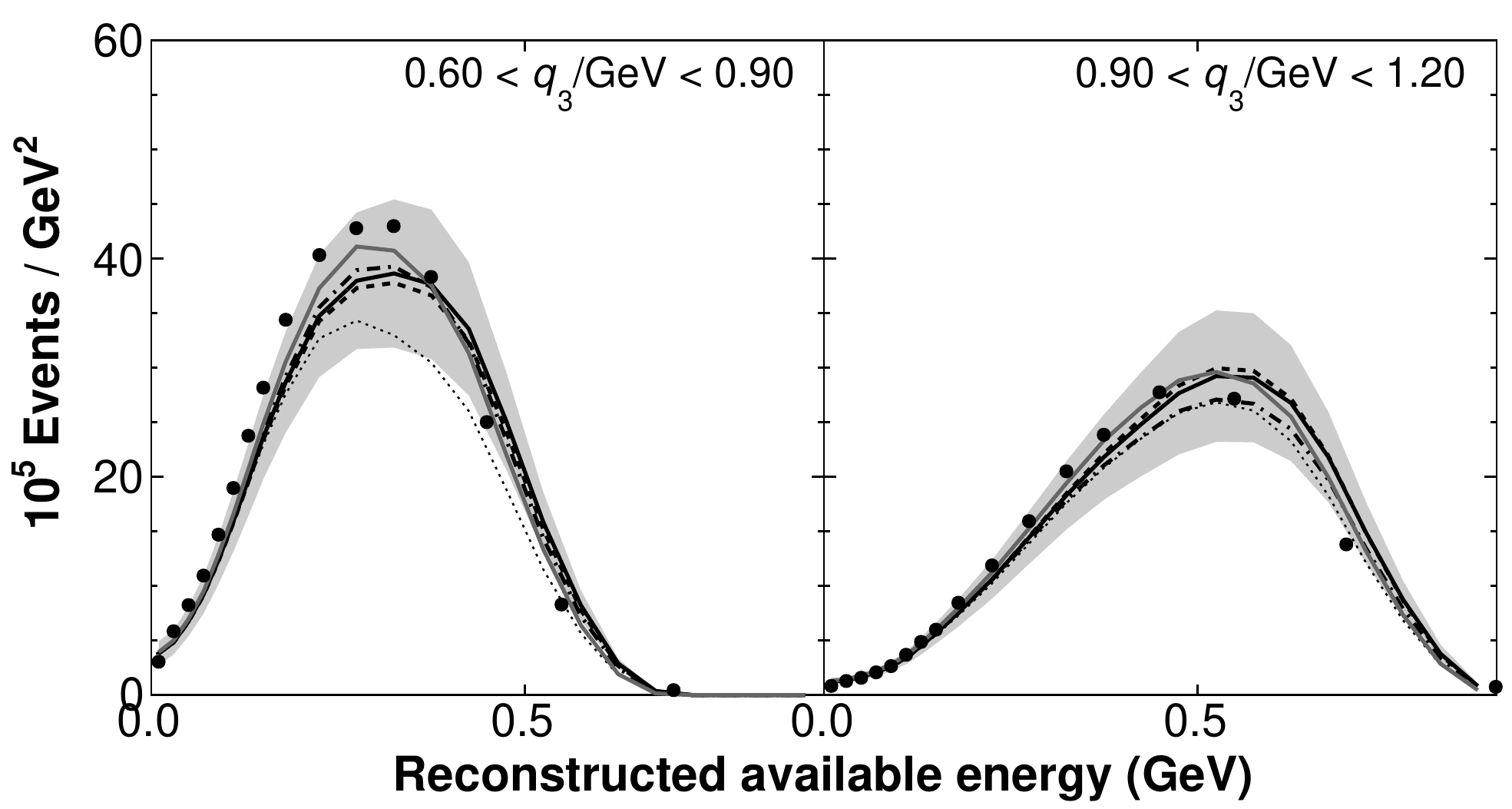}
\caption{Comparison of MnvTune-v1.2 to MnvTune-v3 resonance models designed to reduce the high $E_\text{avail}$ (low $Q^2$) resonance interactions. Points with Statistical Uncertainties only are shown in data. See text for descriptions of each configuration. }
\label{fig:lowRecoilRESstudy}
\end{figure*}

In this section, the reconstructed event distributions are compared to empirical and theory-motivated predictions for the resonance region, with the aim of identifying models that improve the agreement between data and prediction. A suppression of resonance production at low $Q^2$, based empirically on \minerva data is considered, as well as two updates to the Rein-Sehgal model, and a simulation of nucleon removal energy in resonance events similar to the one used in the \genie 2.12 QE process. The results of applying these changes individually are compared to data in Fig.~\ref{fig:lowRecoilRESstudy}.

Prior \minerva results \cite{Eberly:2014mra,McGivern:2016bwh,Altinok:2017xua} have suggested the need for a suppression of resonance production at low $Q^2$ relative to the \genie 2.12 prediction.   A non-linear suppression function from \cite{Stowell:2019zsh} is used, and is 0.4 at $Q^2 = 0$ GeV$^2$ and vanishes by $Q^2 = 0.6$~GeV$^2$.  A similar (but weaker) suppression function has been reported by the MINOS collaboration for interactions on iron at similar energies \cite{Adamson:2014pgc}, and another is implied by the MiniBooNE result in liquid scintillator CH$_2$ below 1~GeV \cite{MiniBooNE:2010eis}. In the current sample, the low-$Q^2$ resonance region is at the highest $E_\text{avail}$ in regions with $q_3 > 0.2$ GeV.  The effect of applying this low-$Q^2$ suppression on top of MnvTune-1.2 is shown in Fig.~\ref{fig:lowRecoilRESstudy} as the low $Q^2\pi$ Supp line (dotted black line); the suppression results in a significant underprediction of the data of the $\Delta$ peak.

One modification to the Rein-Sehgal model in \genie 2.12 combines weighting in $Q^2$ to approximate the Berger-Sehgal model in \genie version 3 and a Pauli-blocking effect.   To produce the $Q^2$ weighting, the ratio of the Berger-Sehgal model in \genie 3.0 to the Rein-Sehgal model in the \genie 2.12 simulation on which MnvTune-1.2 is based is used.  The Berger-Sehgal model \cite{Berger:2007rq} accounts for the lepton mass \cite{Kuzmin:2003ji} and a pion pole term which reproduces the expected Adler-screening effect at low $Q^2$. The implementation \cite{Nowak2009FourMT} and default choice in \genie v3 includes different vector and axial form factor expressions \cite{Graczyk:2007xk} and improved vector form factor parameters determined from the world's (e,e') data \cite{Lalakulich:2006sw};  these combined are the most significant change and also result in a reduced cross section at low $Q^2$. The weight extracted as a function of $Q^2$ from the comparisons of the two versions of \genie yields a maximum low-$Q^2$ suppression of 20\% at $E_\nu=5$~GeV.  Including Pauli-blocking makes the suppression yet stronger.  The overall effect (dashed black line labeled as Pauli B. B-S in Fig.~\ref{fig:lowRecoilRESstudy}) is less than half as strong as the empirical suppression (dotted black line).

Another update of the Rein-Sehgal model has been developed by M.~Kabirnezhad \cite{PhysRevD.97.013002,Kabirnezhad:2017xzx}. It includes non-resonant contributions and the resulting interference terms.  The implementation in {\small NEUT} \cite{Hayato:2009zz} was used to make the ratio to the \genie 2.12 Rein-Sehgal model and to create weights in ($Q^2$, $W$) for the three following channels: \ccpip , \ccnpip , and \ccnpin. The first interesting physics effect is that the total $\Delta$ peak shifts about 20~MeV lower in $W$ due to the interferences.    Secondly, the model's prediction for the non-resonant background rate results in an overall prediction 20\% lower than \genie 2.12 for the higher resonance region.   The net effect is denoted as  the M.K. model line (dash-dotted black line) in Fig.~\ref{fig:lowRecoilRESstudy}.

Finally, the need for fewer events at high available energy and more events in the dip region could be explained by a shift of the $\Delta$ peak in energy transfer and/or  \Eavail (solid gray line labeled as Removal Energy in Fig.~\ref{fig:lowRecoilRESstudy}).    In \genie 2.12, 25~MeV are deducted from the outgoing nucleon for QE interactions to account for its removal from the nuclear potential.  But no such procedure is applied to resonances.   This same shift is implemented for reconstructed resonances:  any charged-current resonance process that produces at least one proton has 25~MeV deducted from \Eavail .  If the only baryons in the final state are neutrons, their energy is not included in the \Eavail estimator and no subtraction is made.   The shift is made to both the true generator quantity and the reconstructed quantity in the simulation.  A study was made with a range of corrections from 0 to 50~MeV in steps of 5~MeV, and 25~MeV provided the best description of this low-$Q^2$ region.   In Fig. \ref{fig:lowRecoilRESstudy}, this special modification is shown. This implementation preserves the total resonance event rate, but shifts it left in each panel, away from the low-$Q^2$ region and toward the dip region.

At face value, this shift accounts only for an oversimulation of the hadronic energy.   Because modifications to the lepton kinematics may be negligible at \minerva energies, it could simplistically account for a real shift in the resonance peak energy transfer.  There is such an effect, long observed in (e,e$’$) data \cite{Bodek:2020wbk,Bodek:2018lmc,Ankowski:2014yfa,Boffi:1993gs,Cooper:1993nx,OConnell:1990njm,Sealock:1989nx,Horikawa:1980cg}, that is stronger at low energy transfers and is also stronger for $\Delta$ than for QE.   Even larger (60~MeV) discrepancies are also observed directly \cite{electronsforneutrinos:2020tbf,Ankowski:2020qbe} running the \genie generator in electron scattering mode.   This intrinsic nuclear effect produces fundamentally different distortions than suppressions due to form factors or Pauli-blocking.   Reality may be a mix of these effects.

\subsection{Model used to produce the cross section}

A single configuration of models is chosen to proceed with unfolding and named MnvTune-v3. We have used the theory-motivated models in place of the empirical ones: the enhancement of the high momentum Bodek-Ritchie tail of the struck nucleon for the QE process, the SuSA~2p2h, and the deduction of 25~MeV removal energy from a subset of resonance reactions.  We retain the RPA effect for QE, and the others at the beginning of this section.  The MnvTune-v1.2 and new models are retained for study of model systematic effects during the unfolding procedure.  The complete baseline model configuration is different from the MnvTune-v1.2 and is also shown in Fig.~\ref{fig:lowRecoilMnv}.

\section{Cross Section Extraction}
The double differential cross-section $d^2\sigma/dE_{\text{avail}}dq_3$ is calculated, using the selected events and subtracting the number of background events predicted by the simulation. 
The background is 1.42\% over the entire sample and consists of neutral current neutrino interactions and antineutrino CC events producing $\mu^+$. 
 
The background-subtracted event sample is unfolded to remove detector resolution effects, using the D'Agostini method \cite{DAgostini:1994fjx} as implemented in RooUnfold \cite{Adye:2011gm}. In order to estimate the validity of the unfolding method, several unfolding studies were done. For each unfolding study, a different model was used as an approximation of the response of the data (as pseudodata),  where Poisson throws were done within the data equivalent statistical uncertainty and unfolded using the MnvTune-v3 model smearing matrix. The number of iterations with which the pseudodata is unfolded is varied, and $\chi^2$ values are calculated by comparing the unfolded pseudodata with the truth pseudodata. In these studies, the $\chi^2$ reached a minimum at two iterations or at ten iterations for some pseudodata variations, including variations with modification of RPA suppression prediction which affects either the low Q2 or high Q2 regions, modification of 2p2h enhancement to include the additional strength being applied fully to nn/pp pairs or np pairs or the high $E_{avail}$ QE tail. We choose a small number of iterations because the large number of iterations causes inflated systematic uncertainties.  Therefore the unfolding is performed with two iterations for this analysis.

\begin{figure*}[ht!]
\centering
\includegraphics[scale=0.65]{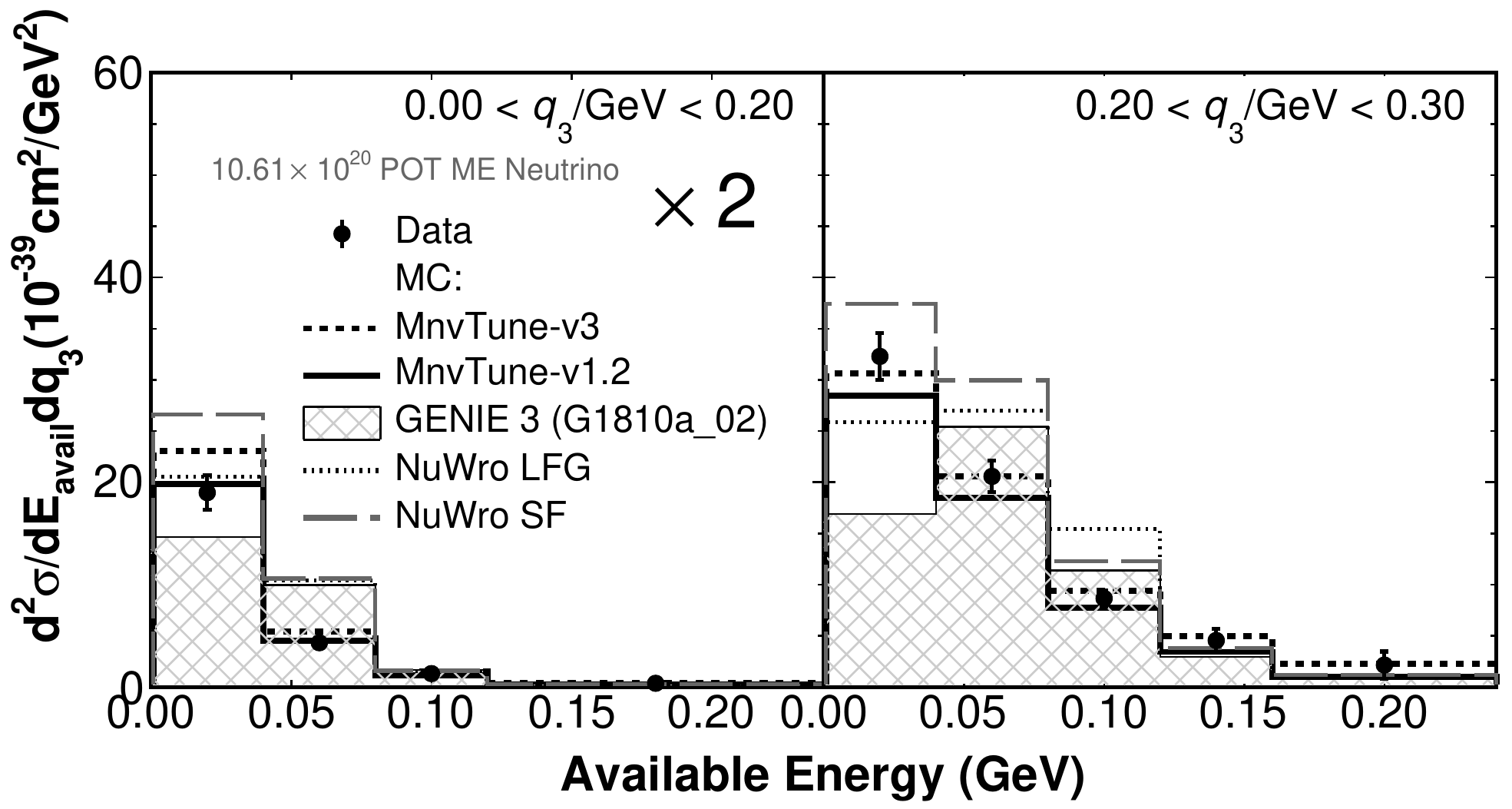}\\
\includegraphics[scale=0.65]{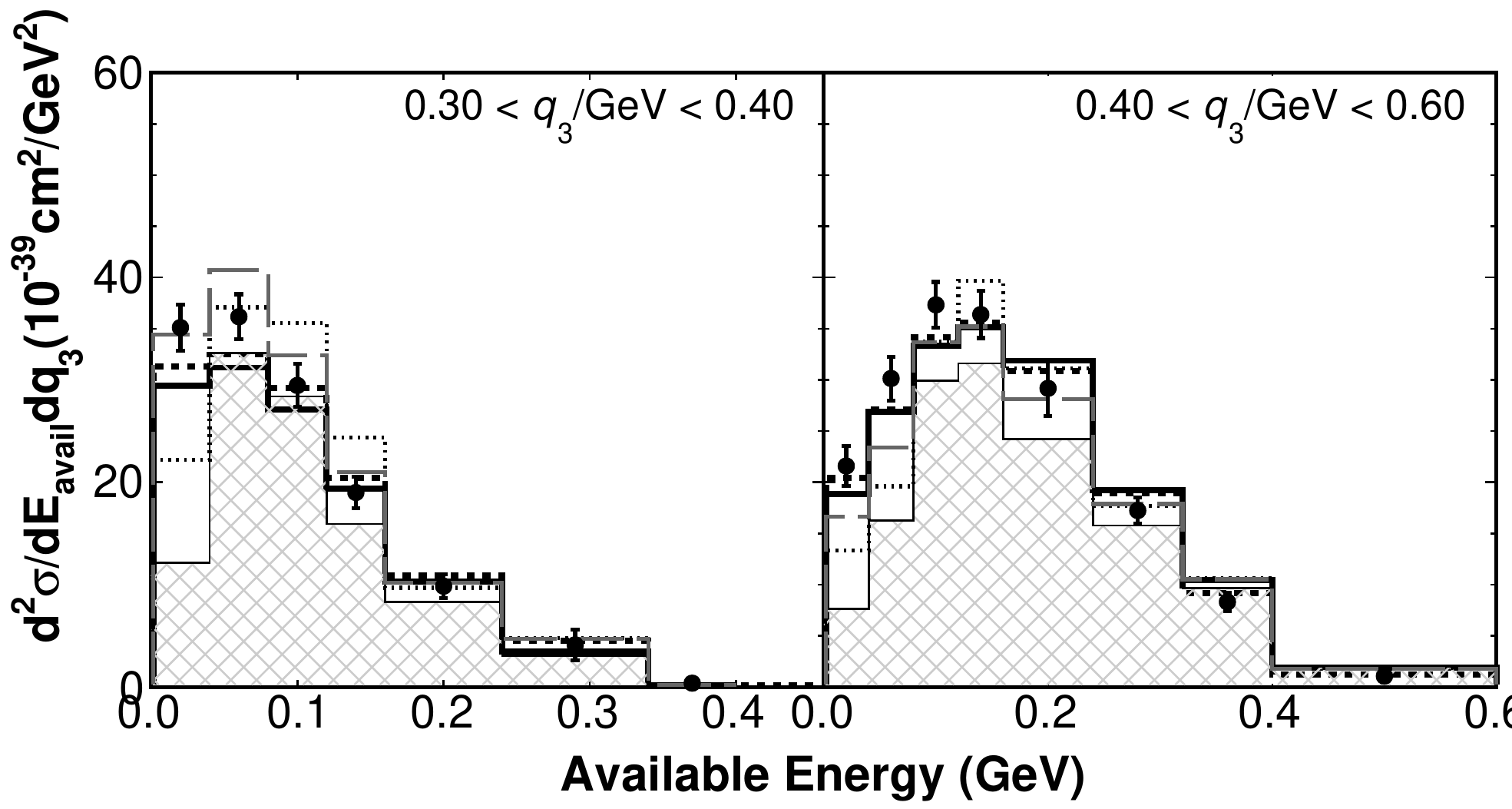}\\
\includegraphics[scale=0.65]{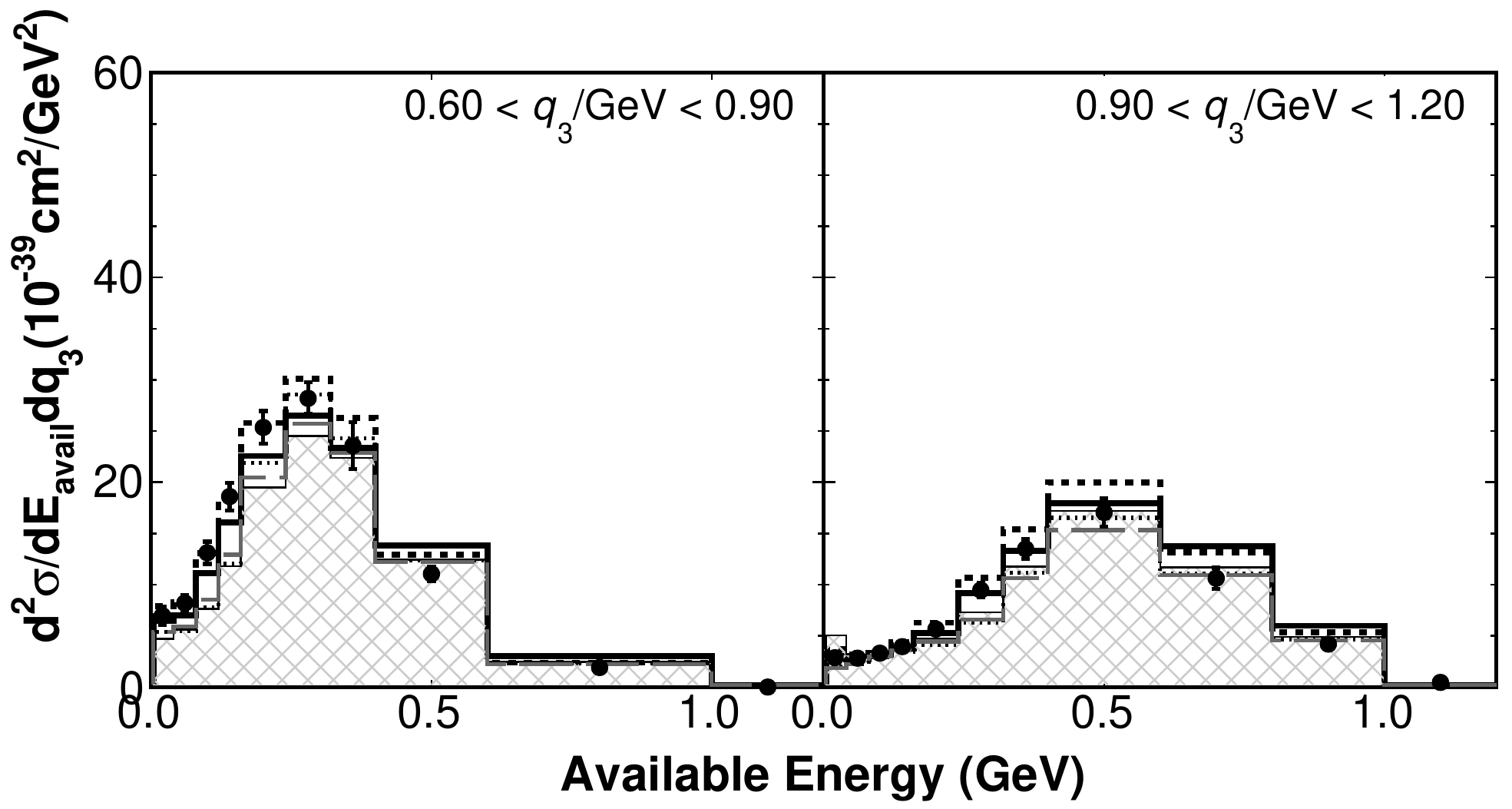}
\caption{Measured double differential $d^2\sigma/dE_{\text{avail}}dq_3$ cross section in available energy and six bins of three momentum transfer is compared to the MnvTune-v3(black dashed line), MvnTune v1.2 (black solid line), NuWro Structure Function (gray solid line), NuWro Local Fermi Gas (dotted gray line)\cite{Stowell:2016jfr}, and GENIE-v3 (gray grid filled histogram). The first $q_3$ panel is scaled by a factor of 2.
}
\label{fig:lowRecoilCrossSection}
\end{figure*}

The number of events as obtained after the unfolding is then divided by the efficiency, which varies from 21\%  to  80\% and is due to a combination of muon acceptance and resolution migration across the $q_3 < 1.2$ GeV analysis boundary, the efficiencies as a function of available energy are reported at the Appendix. The low $q_3$ region bins have, on average, 77\% efficiency; medium $q_3$ region bins, 72\%; and high $q_3$ region bins 53\%, with higher $E_{avail}$ regions having lower efficiency. The normalization factors include 3.115$\times10^{30}$  nucleon targets and the neutrino flux integral from 0 to 100~GeV for an exposure of 1.061$\times 10^{21}$ protons on target.
The measured double differential $d^2\sigma/dE_{\text{avail}}dq_3$ cross section is shown in Fig. \ref{fig:lowRecoilCrossSection}. 

%
%

\section{Systematic Uncertainties}
\begin{figure*}[ht!]
\centering
\includegraphics[scale=0.65]{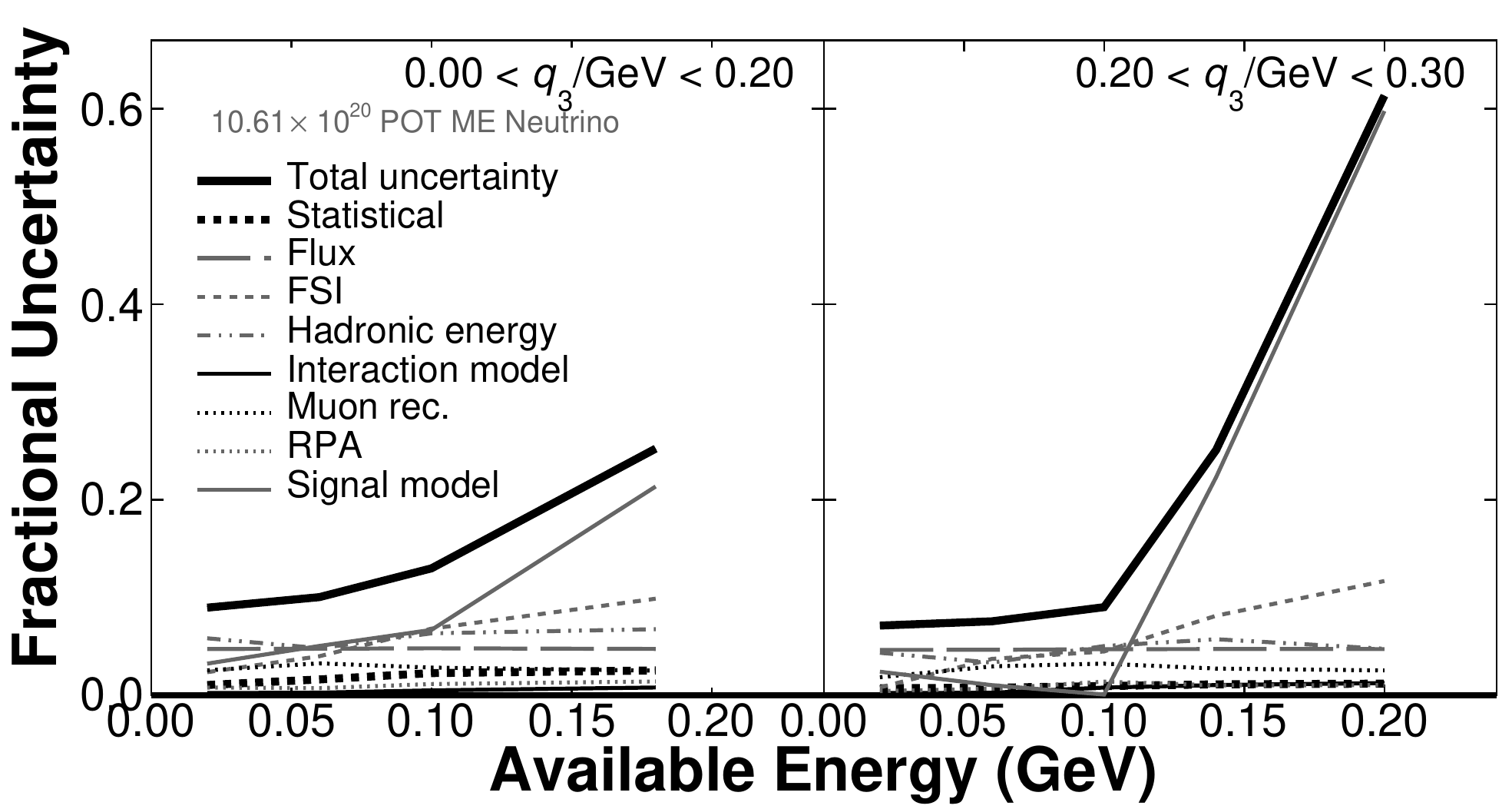}\\
\includegraphics[scale=0.65]{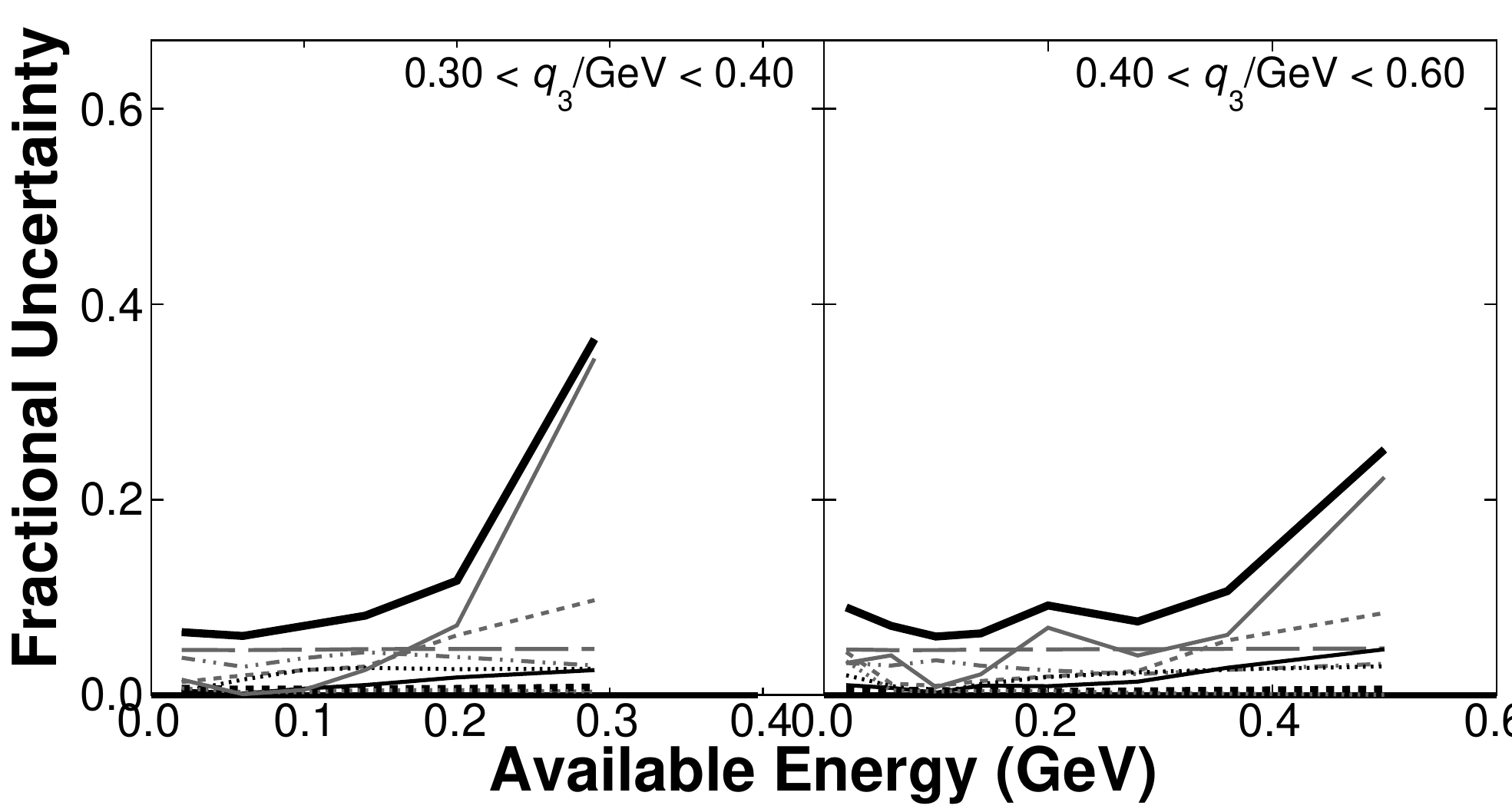}\\
\includegraphics[scale=0.65]{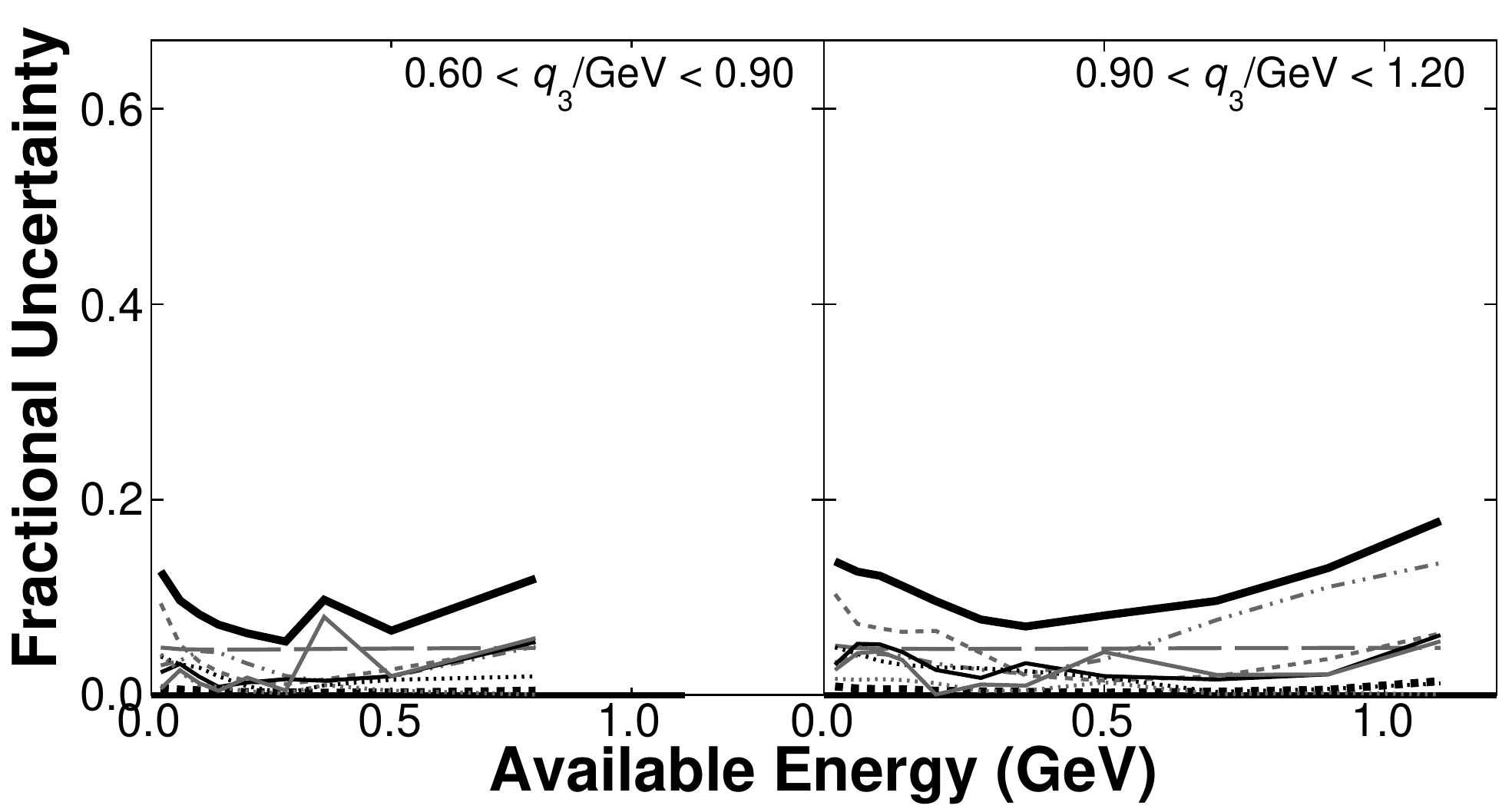}

\caption{Fractional systematic uncertainty breakdown on the double-differential cross-section measurement in slices of $\q$.}
\label{fig:lowRecoilSysUnc}
\end{figure*}

Uncertainties in the measured cross section arise from imperfect knowledge of the flux and detector energy response, uncertainties on the interaction model such as the Final State Interactions (FSI), the QE, 2p2h, and pion event rates, and including uncertainties \minerva assigns as part of modifying the interaction model. 
The breakdown of the fractional systematic uncertainty on $d^2\sigma/dE_{\text{avail}}dq_3$ is shown in Fig.~\ref{fig:lowRecoilSysUnc}. 

To evaluate each source of uncertainty, this analysis re-extracts the cross section using a varied simulation where some parameter has been shifted by its uncertainty. This creates an alternate systematic “universe” whose difference from the nominal cross section can be used to form a covariance matrix.  The covariance matrices for all sources of systematic uncertainty are summed to form a total covariance matrix.  The error bars in Fig. Fig.~\ref{fig:lowRecoilSysUnc} represent the diagonal elements of this matrix.   In some cases, a model parameter can be varied in more than one way (such as shifting an energy scale by plus or minus one standard deviation).  In these cases, the resulting covariance matrices are averaged to form the covariance matrix for that uncertainty.  In the case of the flux uncertainties, which arise from many underlying correlated parameters, hundreds of universes are generated with those parameters randomly varied according to their probability distributions and taking into account correlations.  As in other cases, a covariance matrix is formed for each universe and the universes are averaged to form the flux covariance matrix. 

The flux uncertainty (long dashed gray line in Fig.~\ref{fig:lowRecoilSysUnc}) is practically constant with $E_\text{avail}$ and $q_3$, around 4.7\%.   It is  obtained from comparisons of the hadron production model with data from hadron production experiment NA49 \cite{NA49:2006oyk}, focusing effects, and an in-situ neutrino-electron scattering constraint in the medium energy beam \cite{MINERvA:2019hhc}.

The uncertainty in the detector energy response is divided into two uncertainties,  hadronic energy and muon reconstruction uncertainty (dash-dotted gray and thin dotted black lines respectively in Fig.~\ref{fig:lowRecoilSysUnc}). The hadronic energy uncertainty varies throughout the distribution and rises to 10\% at high $0.9 < q_3 < 1.2$ GeV.  The input uncertainty is determined from hadron calorimetry data taken with a test beam detector \cite{MINERvA:2015yej}. The smaller detector response uncertainty is associated with the muon energy measurement, after a muon scale correction described in \cite{MINERvA:2019gsf,MINERvA:2021mpk}, and the muon angle. 

Uncertainties coming from GENIE event generator are divided into two main groups, FSI and interaction model uncertainties, shown in Fig.~\ref{fig:lowRecoilSysUnc}. The FSI uncertainty has sub-component contributions, where the nucleon and pion mean free paths dominate in the $\q<$ 0.4 GeV region at around 9\%. The higher $q_3$ region has contributions mainly from pion elastic scattering and pion inelastic scattering with a 6\% uncertainty. Thus, the total FSI uncertainty contribution in the cross-section measurement in many bins is around 10\%.  The GENIE interaction uncertainties that govern pion production rate in the resonance region are about 4\% on the extracted cross section.   These are much lower than their effect in the error band in Fig.~\ref{fig:lowRecoilMnv} where they are among the most important.  This is expected because cross section model errors affect the migration matrix less than the absolute rate, and the unfolding iterations reduce the effect of model errors even further.

The model used to extract the cross-section involves numerous modifications to GENIE.  We have built additional uncertainties around these modifications and applied them using the universe method described above.   The $M_A^{QE}$  uncertainty is smaller than and replaces the official GENIE uncertainty (-15\%, +25\%) within the ``Interaction model'' category.  The new uncertainty is approximately $\pm$9\% and is from the analysis in \cite{Meyer:2016oeg} and its effect on the QE region extracted cross section is 3\%.  An uncertainty on the RPA effect is documented in \cite{Gran:2017psn}.  Because of its importance to this sample it is shown separately, but its effect is also small on the unfolded cross section.

Finally, an uncertainty is assigned corresponding to the difference between the cross sections extracted using the MnvTune-v1.2 model and using the new model MnvTune-v3 (Signal model, gray solid line in Fig. ~\ref{fig:lowRecoilSysUnc}). This captures the most significant signal model uncertainty implied by the different 2p2h models and the lowest-$Q^2$ resonance model variations.  The 2p2h and QE components contributions are similar to the axial mass effects noted above.  The choice of low-$Q^2$ resonance modification has the largest effect in the low-$Q^2$ bins.

%
%
\section{Cross Section compared to Generator Predictions}
\label{sec:results}
The resulting double differential cross-section shown in Fig. \ref{fig:lowRecoilCrossSection} is compared  to several neutrino event generator predictions: MnvTune-v3, MnvTune-v1.2, two configurations of NuWro \cite{Golan:2012rfa}, and the GENIE v3.0.6 configuration G18\_10a\_02\_11a. The GENIE v3.0.6 uses a local Fermi gas (LFG) and the Valencia model for both QE with RPA \cite{Nieves:2004wx} and 2p2h \cite{Nieves:2011pp}. Berger-Sehgal replaces Rein-Sehgal \cite{Berger:2007rq} for resonance plus other changes described and cited earlier.  And the Bodek-Yang DIS model \cite{Bodek:2002ps} is still used, including the non-resonant background in the resonance region in this analysis. We chose the empirical hA FSI model for these comparisons to match the one we use with the GENIE v2 generator.   A summary of generator configuration details and complementary comparisons to previously published data can be found in \cite{avanzini2021comparisons}.  The two NuWro configurations presented here use different QE models, a local Fermi-gas with RPA effects in once case and a global Fermi-gas with a spectral function initial nucleon state in the other.   The former is the one used in \cite{avanzini2021comparisons}.   Then NuWro implements only the $\Delta$ resonance, leaving the non-resonant scaled Bodek-Yang DIS model to cover the rest of the rate the resonance region.  The model by Salcedo and Oset \cite{Salcedo:1987md} is used for FSI rescattering of hadrons as they exit the nucleus.  Because the MnvTune-v3 had model elements chosen to describe the reconstructed data better, the overall agreement with the extracted cross section is also good.  The NuWro and \genie v3 models describe the data in some parts of the kinematic space but deviate systematically in the QE and 2p2h rich regions. Because many effects overlap, we need to use the variety of models available to try to disentangle which processes or combinations of processes need additional work. 

The comparisons between data cross section measurement and different model predictions in Fig. \ref{fig:lowRecoilCrossSection}  shows the models are far from perfect in many regions. The full $\chi^2$ with covariance does not have a simple interpretation when comparing the two MnvTunes, both of which are far better than the other three models.  The $\chi^2$ values and commentary are provided in the appendix.

\subsection{The QE process}

The largest variations among the models are in the QE region, roughly the lower half of reported data in each panel of $q_3$.  In fact, the differences between the two versions of NuWro are only in the QE process, and can be used to illustrate what impacts this part of the distribution.

The spectral function model (NuWro SF, dashed gray) is very different largely due to the lack of an RPA screening (or empirically equivalent \cite{Martini:2016eec,Nieves:2017lij}) effect, and not due to the spectral function. It produces higher predictions than the data and all other models in the first one or two bins in panels with $q_3 < 0.4$ GeV.  In the NuWro spectral function, the cross section is integrated using a two-dimensional distribution of initial state off-shell nucleon momentum and removal energy.  Its implementation is fundamentally different from the modified Bodek-Ritchie tail in the MnvTune-v3 (dashed black), though they share theoretical motivation.  Fig.~\ref{fig:lowRecoilMECQEstudy} shows two models with 5\% to 10\% differences in the QE region.  Unlike the NuWro comparison here, both models have the RPA effect, so the spectral function and 2p2h changes are the only ones active, and the spectral function effect on its own has less than 5\% effect.

The local Fermi gas version (NuWro LFG) under-predicts the data in the lowest bin in all panels then becomes similar to the other NuWro model. It does have an RPA screening effect at low $Q^2$.  Its implementation is independent from the Valencia model \cite{Nieves:2004wx} but is intended to produce a similar prediction.  It is also different than the Valencia model RPA weight applied to the two GENIE v2 tunes, and it underpredicts the data in the first bin of all panels. But in other bins at the QE peak and the next several higher, this model is higher than the NuWro SF model and much higher than the GENIE models, even the similar MnvTune-v1.2 and \genie v3, and is a poor description of the data overall.  The eponymous LFG initial nucleon momentum distribution does not include high momentum nucleons and has lower momenta on average than a regular Fermi gas.  This  would produce a narrower QE peak, however its effects are difficult to discern.

A second trend is evident by comparing NuWro and \genie \minerva tunes relative to the data: in the next three to five available energy bins, as a function of $q_3$ NuWro goes from overestimating the data to systematically underestimating it. In contrast, the two modified GENIE models are close to the data and each other (by construction), with some data points favoring one or the other. These bins are a mix of the QE peak, 2p2h, and a little $\Delta$ resonance. Mismodeling the relative strength of these three processes would account for the NuWro behavior.  Even just a form factor effect (such as $M_\text{A}^{\text{QE}}$) for QE could account for some of these differences in the total rate and $q_3$ trends for the generators.

\subsection{The lowest available energy bins}

The predictions diverge the most in the lowest available energy bins in each panel. In addition to the RPA effect within the QE model described in the previous subsection, three other generator design choices have a large impact on the prediction for these bins.  Two come from the strength of the FSI processes that produce neutrons and low energy nucleons in the final state.  One is from the way the nucleon removal energy cost is applied to the resulting hadron system for QE.  These mechanisms are described in turn with estimates of the magnitude and what is available in the neutrino event generators. 

Events where the QE proton turns into a neutron before exiting the nucleus is a special component of these bins. In the first \Eavail bin of the lowest $q_3 < 0.6$ GeV panels, the GENIE v2 tunes predict 20\% of the events are QE events with energy transfers above 100 MeV and feed down via a FSI process, 10\% are 2p2h events and 2\% are resonance events with the same kinematics.  In the extreme case, 10\% of the events have only neutrons in the final state and therefore exactly zero \Eavail .  These happen when the generator’s FSI model produces either the $p \rightarrow pn$ knockout process, the $p \rightarrow n$ charge exchange process, or pion absorption followed by ejection of two or more nucleons. The empirical tuned hA FSI model is nearly the same for all three \genie versions \cite{Harewood:2019rzy} but different from the Oset model \cite{Salcedo:1987md} used by NuWro.  A study of the hA vs. hN models in \genie 3 (not shown) suggests 10\% differences in these bins from this choice alone.

A related generator choice is that FSI can be suppressed, either for convenience (low energy nucleons are not observed in Cherenkov detectors) or via applications of Pauli blocking of the rescattered state.  For example, a 24 MeV proton might never be transformed into a 4 MeV proton and a 20 MeV neutron.  A recent discussion of these effects is presented in \cite{Dytman:2021ohr}; Fig.~5 of that paper shows dramatic differences below 50 MeV proton kinetic energy in generators (along with proton carbon scattering data) from the FSI model configuration alone.  The \genie hA and NuWro models in that paper produce a similar prediction as does the INCL++ based model in a forthcoming version of \genie .  Therefore, this effect should not be present in the comparisons in Fig.~\ref{fig:lowRecoilCrossSection}.   In contrast, the \genie hN model and NEUT \cite{Hayato:2009zz} should deviate significantly in the lowest \Eavail bin.

The \genie v3 generator dramatically underpredicts the first bin of the data.  This comes from a change in how nucleon removal energy is treated.  In \genie v2 (including the two MnvTunes), 25 MeV is explicitly subtracted from the proton for QE reactions.  This 25 MeV is also subtracted from the hadron state in our special resonance modification.  In \genie v3, this subtraction is not made, and the resulting distribution of \Eavail is systematically shifted higher. To reiterate the design, in the classic (e,e$’$) nuclear effect paper \cite{Moniz:1971mt}, the QE peak is higher by 25 MeV.  In \genie this is accomplished by using the deForest prescription \cite{DeForest:1983ahx}.  With Pauli-blocking and the final 25 MeV subtraction, the resulting protons in \genie v2 are produced down to zero kinetic energy. The strategy starts the same in \genie v3, but the Pauli-blocking step is not followed by a 25 MeV subtraction, so very few protons are produced below 20 MeV.  Discussion of additional ways to treat these effects were described in Sec.~\ref{sec:simulation} for resonances and can be found in \cite{Bodek:2020wbk, Bodek:2018lmc} and has been implemented for QE in \cite{MINERvA:2019ope}. 

Collectively these model choices create $\pm$20\% differences in the QE-rich first bin in each panel.  In our extraction of the cross section, we have applied the uncertainty to the RPA effect \cite{Gran:2017psn}.  Its magnitude in the reconstructed distribution is about half the size of differences described in this subsection.  Luckily, it has little impact on the extracted cross section. Recent oscillation experiments have used a combination of RPA, FSI, and removal energy uncertainties.   These uncertainties are justified or would be more than needed if their base model was already close to the \minerva data points.

\subsection{The low Q2 resonances}

All generator configurations in \ref{fig:lowRecoilCrossSection} describe the low-$Q^2$ resonances better than MnvTune-v1.2 and the Rein Sehgal model.  The MnvTune-v3 uses a new prescription to apply a removal energy to resonances similar to what \genie v2 does for the QE process.  It preserves the event rate but shifts it to lower $\Eavail$.   Its effects compared to the \genie v3 model are already described in Sec.~\ref{sec:simulation} and Fig.~\ref{fig:lowRecoilRESstudy} which are a reduction in the event rate compared to MnvTune-v1.2.  The result in the \genie v3 curve in Fig.~\ref{fig:lowRecoilCrossSection} is consistent with the isolated study.   Ours is the first exploration of such a removal energy effect serving analysis of resonance data, but a suppression effect may also be needed to describe these data.

The pion production models in NuWro are within a few percent of the MnvTune-v3 in the bins where this effect is significant.  The $\Delta$ model is from Lalakulich and Paschos \cite{Lalakulich:2005cs} with deuterium-data based axial and vector form factors \cite{Lalakulich:2006sw} and Pauli blocking. It would be similar to the dashed black line in Fig.~\ref{fig:lowRecoilRESstudy}.  The higher resonances are treated very differently in NuWro, preventing firm conclusions.  Instead of simulating the non-$\Delta$ resonances and their decay like other generators, the NuWro event rate is entirely provided by the DIS model using only the quark-hadron duality principle to reproduce the resonance interaction rate on average.

\subsection{Comparison to previous measurement}


This result has several major improvements compared to the original measurement with the Low Energy dataset~\cite{Rodrigues:2015hik}.  These changes cause the reconstructed distributions to differ even with a consistent MC configuration like MnvTune-v1.2. The magnitude of these effects is 10\% to 20\% in some regions of the sample.  The most significant contribution is from the estimated flux; the ME result uses the 12\% $\nu+e$ scattering adjustment~\cite{MINERvA:2019hhc}.   In contrast, the 8\% LE adjustment~\cite{Park:2015eqa}  was not yet available for the first analysis. The +3.6\% muon energy scale correction that is applied to the ME data \cite{MINERvA:2021mpk} has complex effects on this sample and is also significant. There are numerous improvements of 2\% or less including the detector mass model and efficiency corrections.  And the sophistication of the uncertainty budget is improved.

Due to a collection of additional effects, the unfolded data are similar in parts of the sample and different in others.  The unfolded distributions use different central value MC.   The most prominent change is that the original analysis did not have any addition to the 2p2h or QE rate in the dip region, that aspect of MnvTune-v1.2 was added afterward.   Of equal significance, the original analysis used neither a low-$Q^2$ resonance suppression nor a $\Delta$ hadronic energy shift to account for that poorly predicted region of the sample. 

A third effect comes from the unfolding technique, which introduces shifts in the data/MC ratio in Fig.~\ref{fig:lowRecoilCrossSection} compared to the ratio in the reconstructed distributions in  Fig.~\ref{fig:lowRecoilMnv}.  Some of these shifts are from features encoded in the migration matrix, such as how many events migrate into the lowest \Eavail bins.  Other shifts arise from the iterative unfolding method.  When the input model is far from the data, iterations can lessen the gap, removing some of the bias from an imperfect starting model.  In this analysis, iterations move the ratio by 0.08 in the QE region compared to the ratio after the first unfolding and the reconstructed distributions, closing the gap.  The discrepancy started at 25\% and was reduced to 17\% for MnvTune-v1.2, and went from 16\% to 9\% for MnvTune-v3.  In the QE region of the previous measurement, the reconstructed distribution was already well described, so iterations had negligible effect.   

The new results in this paper are the best starting point for interpreting the cross section in the QE, 2p2h, and $\Delta$ regions.  Because of these changes, combining the published unfolded LE result with the new ME result is not a viable analysis, and we consider the previous results superseded by the new results. The changes individually and collectively are more significant than the expected theoretical cross section effects for the two flux-averaged cross sections.   Future analysis to bring the LE cross sections on the same footing or a joint analysis with the reconstructed data using the data preservation packages \cite{Fine:2020snd} may shed more light on the neutrino energy dependence of the conclusions reached by this data set alone.

%
%
\section{Conclusions }
This paper presents an analysis of inclusive charged-current neutrino interactions on a hydrocarbon (polystyrene) target at low three-momentum transfer along with several model variations for these interactions.   The measured double-differential cross section, as a function of three-momentum transfer and available energy, are shown with comparisons to three variations of \genie and two configurations of NuWro event generators.  The QE, 2p2h, and resonance processes are not well described by these generators, though the latest model elements are significant progress.

An analysis of reconstructed data distributions with several model elements is used to produce a new central value (MnvTune-v3) for unfolding and evaluation of systematic uncertainties, compared to prior \minerva measurements.  The SuSA prediction for the 2p2h model, an enhancement of the high momentum tail of the struck nucleon momentum for QE, and a deduction of 25 MeV removal energy from the resonance final state replace empirical tunes to \minerva data.  Model choices with similar effects are available in \genie v3 and NuWro.   

Even with the improved agreement from the MnvTune-v3, discrepancies remain: in the reconstructed distributions much of the QE region is underpredicted by 5\% to 20\%.   As suggested by the reconstructed distribution uncertainty and how the generator predictions differ in Fig.~\ref{fig:lowRecoilCrossSection}, there may be enough model freedom that there are multiple ways to describe the data.  There are several theory-motivated modifications to the QE that can be studied in the future.   The axial form factor and the RPA screening effect are uncertain and the QE rate can be reduced.  More subtle implementations of the removal energy would change the prediction, as would the many beyond-Fermi gas nuclear models. Plus this region has feed-in via final state hadron rescattering with uncertain strength.  The measurement in this paper is a benchmark for future improvements in the modeling needed for precision neutrino oscillation experiments.

%
%
\section*{Acknowledgements}
This document was prepared by members of the \minerva Collaboration using the resources of the Fermi National Accelerator Laboratory (Fermilab), a U.S. Department of Energy, Office of Science, HEP User Facility. Fermilab is managed by Fermi Research Alliance, LLC (FRA), acting under Contract No. DE-AC02-07CH11359.
These resources included support for the \minerva construction project, and support for construction also was granted by the United States National Science Foundation under Award No. PHY-0619727 and by the University of Rochester. Support for participating scientists was provided by NSF and DOE (USA); by CAPES and CNPq (Brazil); by CoNaCyT (Mexico); by Proyecto Basal FB 0821, CONICYT PIA ACT1413, and Fondecyt 3170845 and 11130133 (Chile);  
by CONCYTEC (Consejo Nacional de Ciencia, Tecnolog\'ia e Innovaci\'on Tecnol\'ogica), DGI-PUCP (Direcci\'on de Gesti\'on de la Investigaci\'on  - Pontificia Universidad Cat\'olica del Peru), and VRI-UNI (Vice-Rectorate for Research of National University of Engineering) (Peru); NCN Opus Grant No. 2016/21/B/ST2/01092 (Poland); by Science and Technology Facilities Council (UK); by EU Horizon 2020 Marie Skłodowska-Curie Action; by a Cottrell Postdoctoral Fellowship from the Research Corporation for Scientific Advancement; by an Imperial College London President's PhD Scholarship.  We thank the MINOS Collaboration for use of its near detector data. Finally, we thank the staff of
Fermilab for support of the beam line, the detector, and computing infrastructure.
\bibliography{Analysis}

\newpage
\appendix
\section*{Appendix}

\begin{figure}[ht!]
\centering
\includegraphics[scale=0.35]{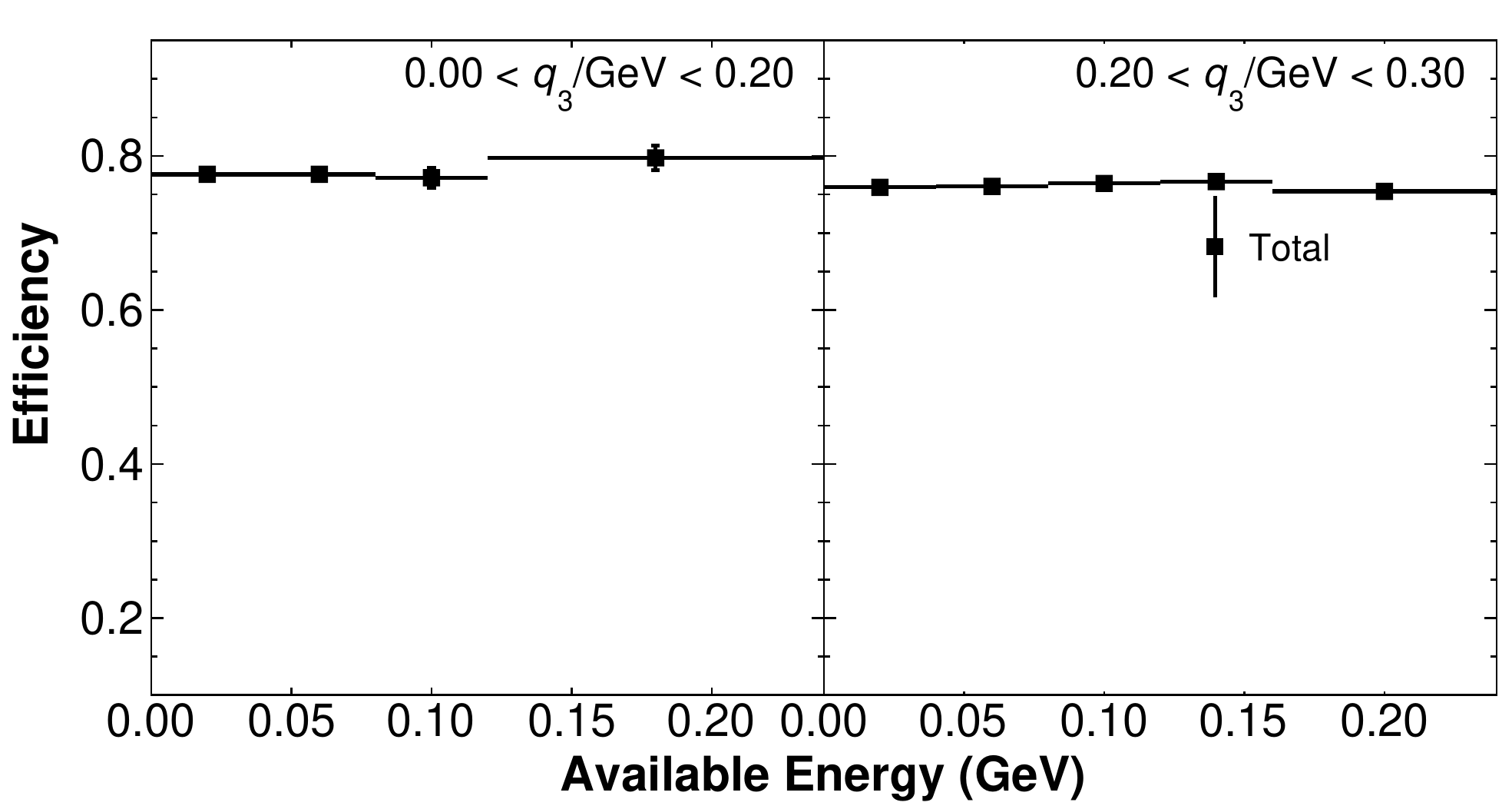}\\
\includegraphics[scale=0.35]{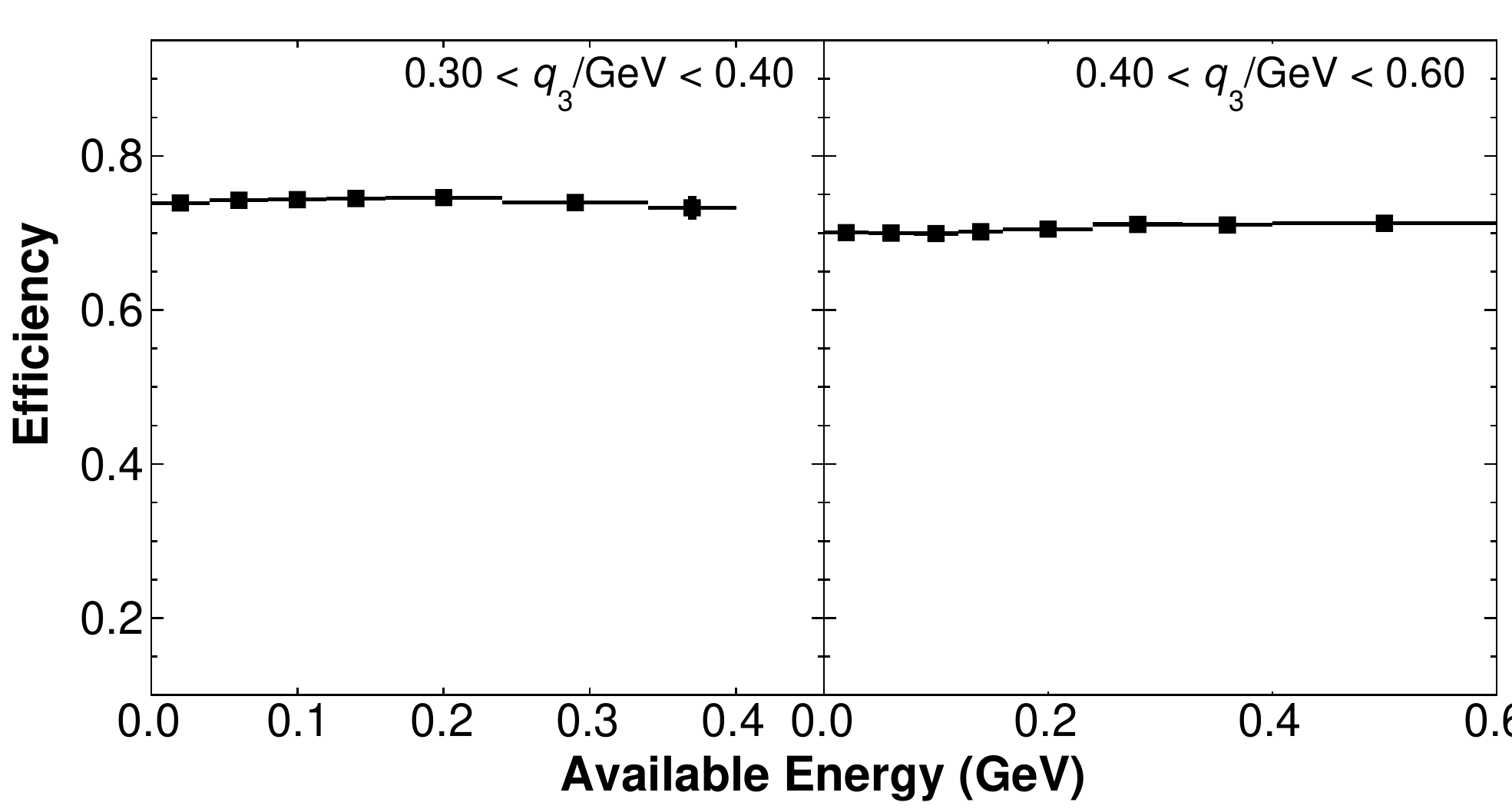}\\
\includegraphics[scale=0.35]{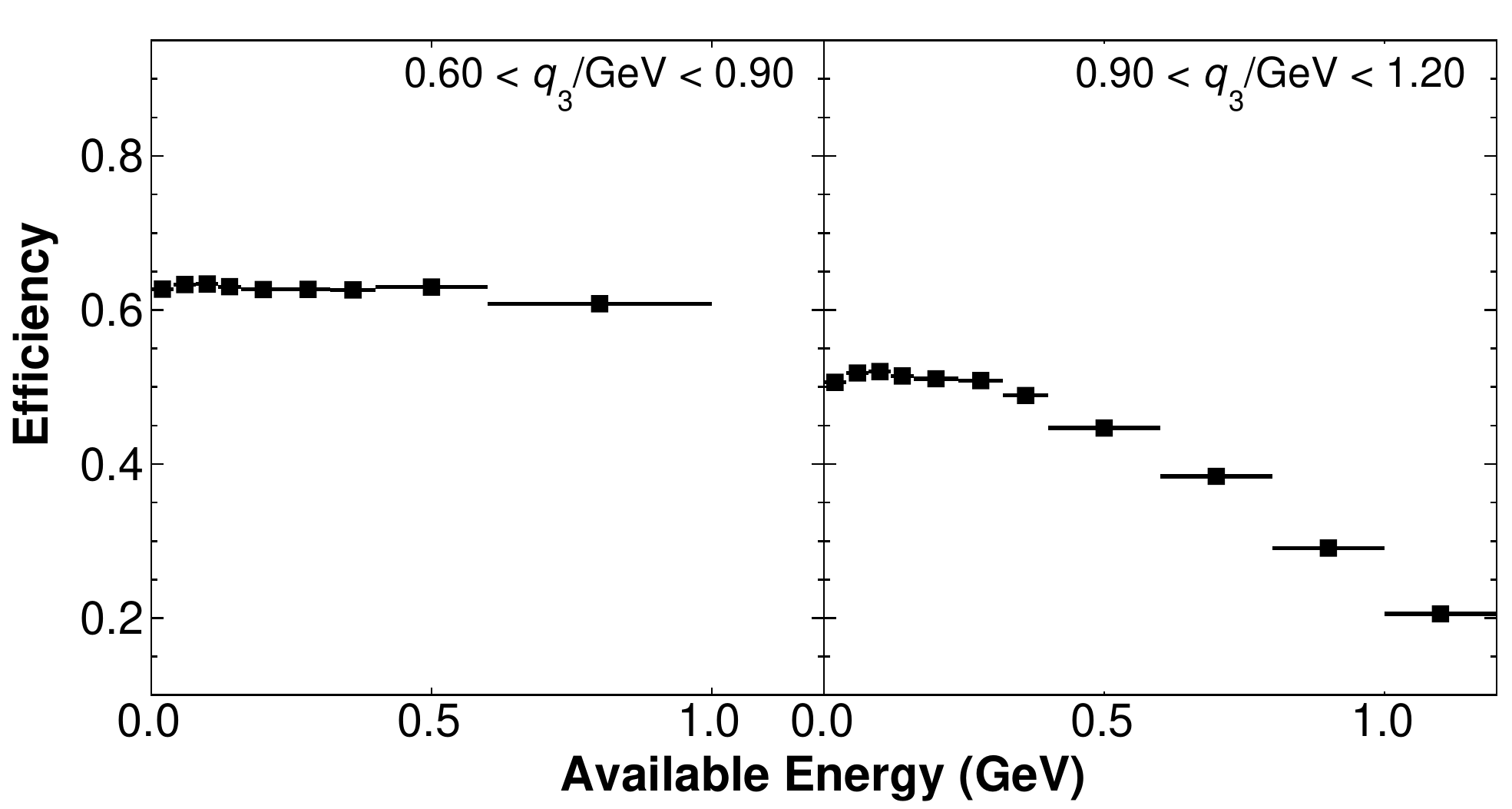}
\caption{Total efficiency as a function of Available Energy used in the double-differential cross-section measurement in slices of $q_3$}
\label{fig:lowRecoilEff}
\end{figure}

To produce the cross section, the samples are corrected for an efficiency which is estimated by the MC.  The net effect is shown in Fig.~\ref{fig:lowRecoilEff}. The efficiency is high overall because the analysis is inclusive within the kinematic region of the analysis.  The efficiency accounts for a wide range of effects, including an event being reconstructed in the fiducial volume, a random contribution to having a muon matched to the MINOS muon spectrometer, being selected as a negative muon, and having an event rejected due to other reconstruction artifacts such as unrelated beam activity.  The downward trend in the last panel is a combination fluctuations out of the region to higher $q_3$ and the systematically lower efficiency to match muons near the 20 degree boundary used in the analysis. 

The following table summarizes the five cross section models in terms of their $\chi^2$ with 44 degrees of freedom.  The $\chi^2$ for the two MnvTune configurations are close to each other and much better than the other three models.  With the full covariance, MnvTune-V1.2 has the lowest chisquare.  With the diagonal elements only MnvTune-V3 is lower, 79 vs. 140 for 44 degrees of freedom. 

\begin{table} [H]
  \centering 
  \begin{threeparttable}
    \begin{tabular}{ccc}
    MC/Generators  & $\chi^2$ & $\chi^2$/NDF\\
     \midrule\midrule
      MnvTune-V3  &   1100.8   &   25.       \\
    \cmidrule(l  r ){1-3}
     MnvTune-V1.2  &   963.2   &   21.9       \\
    \cmidrule(l r ){1-3}
     NuWro SF & 9981.8 & 226.9 \\ 
    \cmidrule(l r ){1-3}
     NuWro LFG & 16363.8 & 371.9 \\ 
    \cmidrule(l r ){1-3}
    GENIE 3 (G18\_10a\_02) & 14148.9 & 321.6\\
    \midrule\midrule
    \end{tabular}
\end{threeparttable}
  \caption{$\chi^2$ of different models compared to data at cross section level using standard $\chi^2$ where there are 44 degrees of freedom. 
}
  \end{table}

\begin{figure}[ht!]
\centering
\includegraphics[scale=0.35]{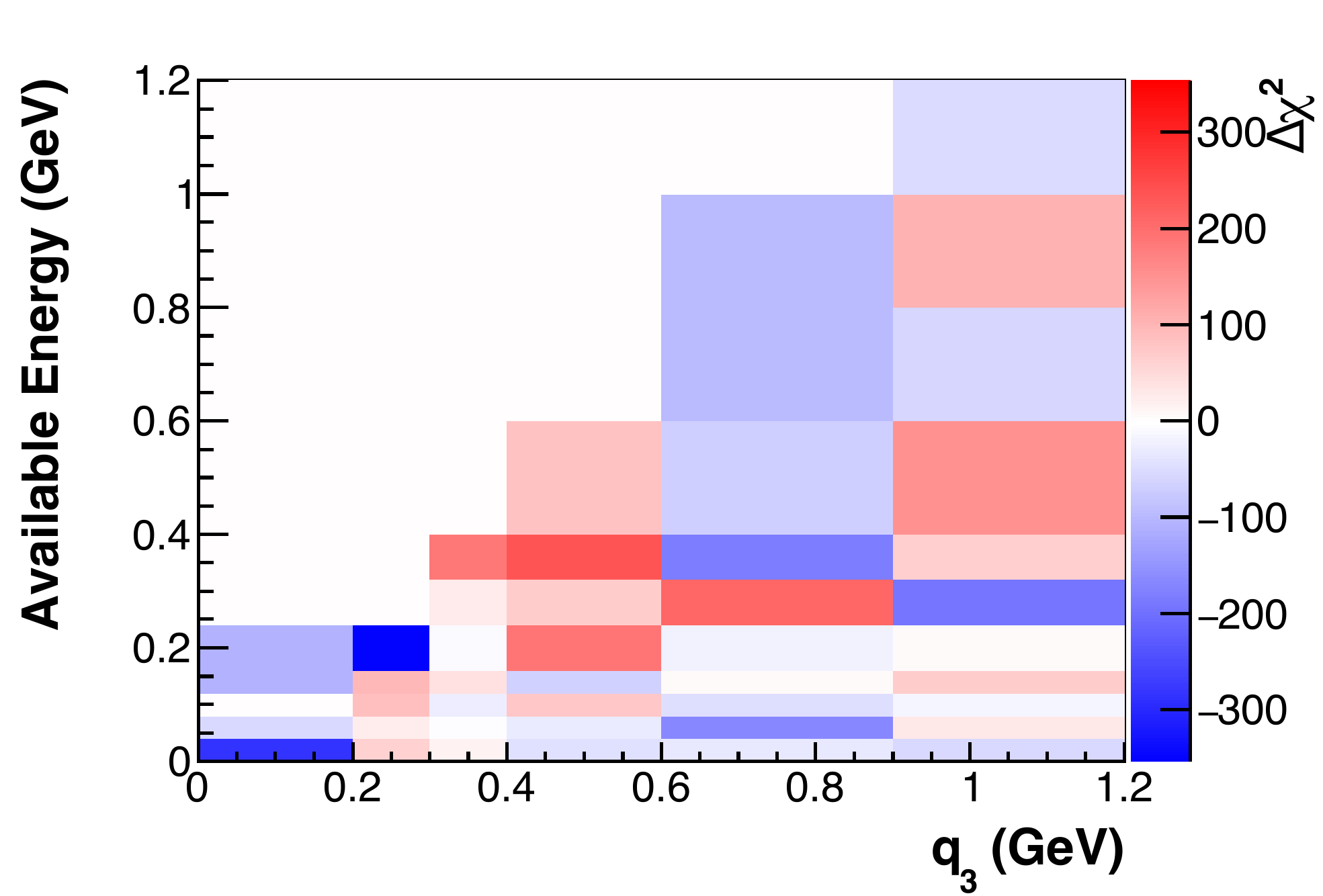}
\caption{Difference of $\chi^2$ between MINERvA Tune v1.2 and MINERvA Tune v3. The negative values suggest MnvTune-v1.2 is better, the positive values favor MnvTune-v3.}
\label{fig:lowRecoilDeltaChi2}
\end{figure}

Another metric used to compare models uses a bin-by-bin $\Delta\chi^2$. Using this metric some illustration of difference aspects of models and where theory motivated modifications indicate improvement and other where the full covariance treatment of the uncertainties supersedes the by-eye comparison in Fig.~\ref{fig:lowRecoilDeltaChi2}.

The bin-by-bin $\Delta \chi^2$ in each $q_3$ and $E_{\text{avial}}$ bin is shown in Fig. \ref{fig:lowRecoilDeltaChi2} and defined as:
\begin{equation}
    \Delta\chi^2_i=\sum_j\bigg(\chi^2_{i,j_{\text{model}}}  - \chi^2_{i,j_{\text{MINERvA Tune v3}}}  \bigg),
\end{equation}
where 
\begin{equation}
    \chi^2_{i,j_{\text{model}}} = \bigg(x_{i,\text{measured}} - x_{i,\text{e}_{\text{model}}} \bigg)\times V^{-1}_{ij}\times \bigg(x_{j,\text{measured}} - x_{j,\text{e}_{\text{model}}} \bigg).
\end{equation}
The i and j indices run over all 44 reported data points. 

The $x_{i,\text{measured}}$ and $x_{i,\text{e}_{\text{model}}}$ represent the cross section measured and expected for the different models respectively \cite{PhysRevD.104.092007}. $V_{ij}$ is the covariance matrix. A negative $\Delta\chi^2$ represents a bin where MnvTune-v1.2 predicts the data better than MnvTune-v3 while a positive value means MnvTune-v3 predicts the data better.


The single-bin contributions vary by hundreds of $\chi^2$ units and don’t obviously correspond to particular components of the physics model or systematic uncertainties. The sum of the bin-by-bin $\Delta\chi^2$ is only 138 which corresponds to the difference in the total $\chi^2$ in Table 1.

Finally, the by-eye interpretation of Fig~\ref{fig:lowRecoilDeltaChi2} and the diagonal element $\chi^2$ favors MnvTune-v3, and there is an evident pattern.  If a single systematic correlates in shape with one of the several discrepancies between model and data, the full covariance $\chi^2$ would naturally discount that particular discrepancy, and a visibly worse model might end with the better $\chi^2$.  The method to study such an effect was to turn off a single systematic and inspect figures equivalent to Fig.~\ref{fig:lowRecoilDeltaChi2} to quantify the most significant changes.  The uncertainty associated with the difference between MnvTune-v3 and MnvTune-v1.2 has by far the largest effect.  The changing contribution to the $\chi^2$ (not shown) is localized to the low-$Q^2$ $\Delta$ production region around $q_3 = 0.5$ GeV and high available energy.  This indicates the systematic associated with the removal energy shift minimizes the effect of that region's discrepancy to increase the MnvTune-v1.2 $\chi^2$. This conclusion reinforces the idea that the two MnvTunes offer similar description of the data overall and that quoting partial (bib-to-bin) $\chi^2$ can help in identifying regions where the data/model tension is more or less pronounced

\end{document}